\documentclass[trackchanges,twocolumn]{aastex701}
\usepackage[english]{babel}
\usepackage[utf8]{inputenc}
\usepackage{amsmath}
\usepackage{amsfonts}
\usepackage{amsthm}
\usepackage{amssymb}
\usepackage{graphicx}

\theoremstyle{definition}

\theoremstyle{remark}

\newcommand{\A}{\mathcal{A}}

\def\actaa{\ref@jnl{Acta Astron.}}      
\def\apjl{\textrm{ApJL}}                
\def\apjs{\textrm{ApJS}}               
\def\ao{\textrm{Appl.~Opt.}}           
\def\aap{\textrm{A\&A}}                

\usepackage[normalem]{ulem}

\begin{document}

\title{Transitions in the Mass-ratio and Spin Properties of Binary Black Holes in GWTC-5}

\author{Elizabeth Flanagan}
\affiliation{Gravity Exploration Institute, School of Physics and Astronomy, Cardiff University, 5 The Parade, Cardiff, CF24 3AA, United Kingdom}
\email{FlanaganE1@cardiff.ac.uk}

\author{Fabio Antonini}
\affiliation{Gravity Exploration Institute, School of Physics and Astronomy, Cardiff University, 5 The Parade, Cardiff, CF24 3AA, United Kingdom}
\email{AntoniniF@cardiff.ac.uk}

\author{Thomas Callister}
\affiliation{Williams College, Williamstown, MA 01267, USA}
\email{tac7@williams.edu}

\author{Debatri Chattopadhyay}
\affiliation{Center for Interdisciplinary Exploration and Research in Astrophysics (CIERA) and Department of Physics \& Astronomy, Northwestern University, 1800 Sherman Ave, Evanston, IL 60201, USA}
\email{debatri.chattopadhyay@northwestern.edu}

\author{Fani Dosopoulou}
\affiliation{Gravity Exploration Institute, School of Physics and Astronomy, Cardiff University, 5 The Parade, Cardiff, CF24 3AA, United Kingdom}
\email{DosopoulouF@cardiff.ac.uk}

\author{Isobel Romero-Shaw}
\affiliation{Gravity Exploration Institute, School of Physics and Astronomy, Cardiff University, 5 The Parade, Cardiff, CF24 3AA, United Kingdom}
\email{Romero-ShawI@cardiff.ac.uk}

\author{Jakob Stegmann}
\affiliation{Max Planck Institute for Astrophysics, Karl-Schwarzschild-Str. 1, 85748 Garching, Germany}
\email{jstegmann@MPA-Garching.MPG.DE}

\begin{abstract}
We analyze the mass-ratio and effective-spin ($\chi_{\rm eff}$) distributions of binary black hole mergers in the latest gravitational-wave catalog, GWTC-5, as a function of primary mass. Using hierarchical Bayesian inference with flexible Gaussian-process population models, we identify four distinct mass regions separated by  transitions in both mass-ratio and spin properties.
Below $\sim18,M_{\odot}$, the population strongly favors equal-mass binaries and exhibits a narrow, symmetric $\chi_{\rm eff}$ distribution peaked at positive values. The positve skewness previously reported for the overall population is absent from this low-mass component and it is most likely  associated with higher mass binaries.
In the range $18$--$30M_{\odot}$, the mass-ratio distribution becomes substantially flatter, while the $\chi_{\rm eff}$ distribution broadens and shifts to a peak consistent with zero. 
 The region associated with the feature near $\simeq35~M_{\odot}$  returns to a narrow $\chi_{\rm eff}$ distribution consistent with symmetry at zero and strongly favors equal-mass binaries. Above $\simeq 45~M_{\odot}$, both the mass-ratio and $\chi_{\rm eff}$ distributions broaden significantly. The inferred support of the spin distribution converges toward the range expected for binaries containing remnants of previous black hole mergers, making the highest-mass region fully consistent with a star cluster population of hierarchical mergers. The close correspondence between transitions in mass ratio and effective spin suggests that different primary-mass ranges trace distinct formation channels, with isolated binary or triple evolution likely dominating the lower-mass population and dynamical assembly becoming increasingly important at higher masses.
\end{abstract}


\section{\label{sec:Intro} Introduction}
The latest gravitational-wave catalogue, GWTC-5 \citep{2026arXiv260527223T, theligoscientificcollaboration2026gwtc50methodsidentifyingcharacterizing}, has significantly expanded the sample of observed binary black hole (BBH) mergers \citep{2019PhRvX...9c1040A,2021PhRvX..11b1053A,LVK_GWTC3_2023,2026arXiv260527223T,2026arXiv260527225T,2026arXiv260527226T}, enabling increasingly detailed studies of the underlying black hole population \citep{2026arXiv260527226T}. While early analyses focused primarily on constraining the overall mass and spin distributions, the growing number of detections has begun to reveal evidence for additional structure within the population \citep{2021ApJ...913L...7A,2019ApJ...882L..24A,Abbott:2020gyp,2026arXiv260527226T,2023PhRvX..13a1048A}. In particular, several studies have suggested the presence of multiple subpopulations distinguished by their masses, spins, and mass ratios \citep{2023arXiv230401288G, 12025arXiv250602250S, 2022ApJ...941L..39W, 2022ApJ...932L..19B,2021ApJ...922L...5C,Antoninietal2023, hussain2024hintsspinmagnitudecorrelationsrapidly, ray2024searchingbinaryblackhole, banagiri2025evidencesubpopulationsmergingbinary, padhyegurjar2026detectionsubpopulationsdelaytimedistribution}. Features such as the excess of high-mass mergers, the existence of systems in or near the pair-instability mass gap, and correlations between spin and mass all point toward the possibility that BBHs form through multiple astrophysical channels \citep{antonini2026a, tong2026a, 2020MNRAS.497.1043D,2022ApJ...928...75H, 2025PhRvL.134a1401A, 2025arXiv250819208M, 2023MNRAS.523.4539K, afroz2025phasespacebinaryblack, afroz2025binaryblackholephase, padhyegurjar2026bbhgenesisdisentanglingbinaryblack}.

Black hole spins provide a powerful diagnostic of binary formation channels \citep[e.g.,][]{biscoveanu2026decadegravitationalwavemeasurementsblack}.
The best-measured spin combination in gravitational-wave observations is the effective  spin parameter~\citep{PhysRevLett.106.241101},
\begin{equation}
\chi_{\rm eff} =
\frac{m_1 \chi_1 \cos\theta_1 + m_2 \chi_2 \cos\theta_2}{m_1+m_2},
\end{equation}
where $m_1$ and $m_2$ are the primary and secondary component black-hole masses, $\chi_1$ and $\chi_2$ are their dimensionless spin magnitudes, and $\theta_1$ and $\theta_2$ are the tilt angles between the individual black-hole spins and the orbital angular momentum. 
Different formation scenarios are expected to produce distinct signatures in both effective spin and mass ratio. In isolated binary evolution, tidal interactions and binary mass transfer are generally expected to align the spins of the black holes with the orbital angular momentum \citep{1981A&A....99..126H, Gerosa_2018, Rodriguez2016c}, leading to preferentially positive values of $\chi_{\rm eff}$.
These channels also tend to favor binaries with nearly equal masses due to the coupled evolution of the stellar progenitors \citep{Marchant2016, 2016MNRAS.460.3545D, 2023ApJ...958...13A}. By contrast, binaries assembled dynamically in dense stellar environments such as globular clusters or nuclear star clusters are expected to have nearly isotropic spin orientations, producing $\chi_{\rm eff}$ distributions centered  at zero and symmetric, extending to negative values \citep{ 2022ApJ...935L..26F, Rodriguez2015a}. Dynamical channels can additionally generate a broader range of mass ratios when hierarchical mergers contribute to the population \citep[e.g.,][]{OLeary2006,Rodriguez2015a,Antoninietal2023}. In such systems, merger remnants from previous generations can pair with lower-mass black holes, flattening the mass-ratio distribution and producing more asymmetric binaries together with larger black hole masses and higher spins \citep{2020ApJ...900..177K,2024arXiv240114837T}.

Additional formation channels may also contribute to the observed BBH population. In hierarchical triple systems, Lidov--Kozai oscillations can efficiently drive binaries to merger while generating substantial spin-orbit misalignment \citep{Silsbee2017,2017ApJ...841...77A, 2021MNRAS.502.2049L, 2019MNRAS.486.4781F,2022arXiv220316544S}. These channels naturally produce $\chi_{\rm eff}$ distributions concentrated near zero but with substantial support extending to both positive and negative values. Triple-mediated mergers are  expected 
to yield a mass ratio distribution peaked at $\sim 1$ but with a significant tail at lower values 
\citep{2022ApJ...937...78M}.
Mergers occurring within active galactic nucleus (AGN) disks provide another possible formation pathway \citep{2024MNRAS.531.3479M, Bartos2016, 2021ApJ...908..194T}. Gas torques and migration traps inside the disk can promote repeated mergers and partial spin alignment, potentially producing systems with large masses, high spins, and asymmetric mass ratios \citep{2021ApJ...908..194T, 2019PhRvL.123r1101Y}. Finally, primordial black holes constitute a qualitatively different scenario in which BBHs form in the early Universe rather than through stellar evolution \citep{2016PhRvL.116t1301B,2016PhRvL.117f1101S}. Depending on the primordial clustering and merger history, these models can produce broad mass-ratio distributions and generically predict very small intrinsic spins, leading to $\chi_{\rm eff}$ distributions sharply peaked near zero \citep{2020JCAP...04..052D,2022PhRvD.105h3526F}. However, recent models show that primordial black holes might also achieve large spins
\citep{2025arXiv250809965D}.

Recent analyses have found growing evidence for such correlations, and, in particular, that the effective spin distribution evolves with black hole mass \citep{2020ApJ...894..129S,2021ApJ...922L...5C, 2022ApJ...928...75H, 2024arXiv240601679P,2025arXiv251105316T,2025PhRvD.111f1305H}. Moreover, binaries with a primary mass $m_1\gtrsim 45M_\odot$ have been shown to have  a broader  $\chi_{\rm eff}$ distribution than lower-mass systems \citep{2025PhRvL.134a1401A,nxnr-pdyx,2025arXiv250602250S,antonini2026a,2025arXiv250915646B,2026arXiv260317987R,2026PhRvD.113j3021M,plunkett2026signaturessubpopulationhierarchicalmergers}. This behavior is consistent with an increasing contribution from hierarchical mergers formed in dense clusters  and  indicates the emergence of a distinct high-mass population. Other studies have similarly argued that the observed BBH population cannot be fully described by a single smooth spin distribution \citep{Wang:2022gnx,2023arXiv230302973L,antonini2026a, 2021ApJ...915L..35K, 2023PhRvD.108j3009G}. 

At the same time, 
 correlations between the mass-ratio distribution, primary mass, and spins have begun to emerge in recent analyses \citep{2021ApJ...922L...5C,2023ApJ...958...13A,2025arXiv250915646B,2025arXiv250818083T,2026arXiv260527226T}. 
 \cite{2026arXiv260527226T} finds that 
  the effective inspiral spin distribution is broader for
unequal-mass binaries and  likely broadens with increasing redshift.
 Measuring how the spin and mass-ratio distributions evolve jointly with primary mass may  provide a direct way to identify transitions between different BBH formation channels and determine whether the observed BBH population is composed of multiple astrophysical components rather than a single smoothly evolving distribution \citep{2020PhRvD.102d3002B,2024ApJ...960...65S}.

In this work, we investigate how the effective-spin and mass-ratio distributions evolve with primary black hole mass using the GWTC-5 catalogue.  Our analysis combines flexible non-parametric Gaussian-process models with simpler parametric descriptions, allowing us to identify which features are robustly supported by the data and which depend on modeling assumptions. In particular, we test whether the  data favor the emergence of distinct subpopulations with different spin and mass-ratio properties at different masses. The increased number of detections in GWTC-5 enables significantly tighter constraints on the possible contribution of hierarchical mergers to the high-mass BBH population and allows us to search for evidence of population transitions as a function of mass and make a stronger connection to astrophysical formation scenarios.
We identify four mass regions across which the mass-ratio and spin distributions change, with transitions at approximately $18$, $35$, and $45,M_{\odot}$. This behavior is consistent with a scenario in which isolated binary or triple evolution dominates the lowest-mass systems and contributes  to most of the  BBH population, while formation in dense environments, such as stellar clusters, becomes progressively more important at higher masses.

\section{\label{sec:Methods} Methods}

With the most recent gravitational wave catalog, GWTC-5 \citep{2026arXiv260527225T} and previous catalogues \citep{Abbott_2023, theligoscientificcollaboration2025opendataligovirgo, RICHABBOTT2021100658}, we infer properties of the black hole population using hierarchical Bayesian inference. Detector improvements in GWTC-5 allowed for these new detected events \citep{2020PhRvD.102f2003B, 2020LRR....23....3A, 2019PhRvL.123w1107T, 2016PhRvD..93k2004M, 2015CQGra..32b4001A, 2015CQGra..32g4001L, 2021PTEP.2021eA101A}. Consistent with previous studies~\cite[e.g.,][]{2025arXiv250818083T}, we focus on sources with a false alarm rate of FAR $< 1$ yr$^{-1}$. Filtering by this rate leaves 259 events. Selection effects are accounted for using recovered injections from injection campaigns \citep{PhysRevX.13.041039, 44x3-hv3y}.

We factorize the astrophysical merger-rate density as
\begin{eqnarray}
\mathcal{R}(m_1,m_2,z,\chi_{\rm eff})
=
\mathcal{R}_{\rm ref}
\frac{f(m_1)}{f(20\,M_\odot)}
\frac{(1+z)^\kappa}{(1.2)^\kappa}\nonumber \\
p(m_2|m_1)
p(\chi_{\rm eff}|m_1).
\end{eqnarray}
Here, \(\mathcal{R}_{\rm ref}\) is the source-frame merger-rate density per unit primary mass at
\(m_1=20\,M_\odot\) and \(z=0.2\), \(m_2\) is the secondary mass, and \(\kappa\)
controls the redshift evolution of the intrinsic volumetric merger rate. The factor
\((1.2)^\kappa\) normalizes the redshift evolution to unity at the reference redshift
\(z=0.2\).

The primary mass spectrum is modeled non-parametrically using a Gaussian process (hereafter $\mathcal{GP}$). We assume a zero-mean $\mathcal{GP}$ with a squared-exponential covariance kernel, such that
\begin{equation}
    f(m_{1}) = \exp[\Phi(\ln m_{1})],
\end{equation}
with
\begin{equation}
    \Phi(\ln m_{1})
    \sim
    \mathcal{GP}
    \big(
    0,\,
    k(x,x^{\prime};a_{m_{1}},l_{m_{1}})
    \big)
\end{equation}
and
\begin{equation}
    k(x,x^{\prime};a_{m_{1}},l_{m_{1}}) =
        a_{m_1}^2 \mathrm{exp}\left[-\frac{(x-x')^2}{2 l_{m_1}^2} \right].
\label{eq:kernel}
\end{equation}
The $\mathcal{GP}$ is evaluated on a log-uniform grid in primary mass over the range $2$--$200\,M_{\odot}$.

{ We similarly model the mass--ratio distribution conditionally on the primary mass as
\begin{equation}
    p(q\mid m_{1}) \propto q^{\beta_{q}(m_{1})},
\end{equation}
where the power-law index $\beta_q$ is itself allowed to vary with primary mass according to a $\mathcal{GP}$ model,}
\begin{equation}
    \beta_q(m_1)=\Xi(\ln m_{1}),
\end{equation}
with
\begin{equation}
    \Xi(\ln m_{1})
    \sim
    \mathcal{GP}
    \big(
    0,\,
    k(x,x^{\prime};a_{\beta_q},l_{\beta_q})
    \big).
\end{equation}
As above, the $\mathcal{GP}$ is constructed with a zero mean and squared-exponential covariance kernel and is evaluated on a log-uniform grid in primary mass between $2$ and $200\,M_{\odot}$.

We model the $\chi_{\rm eff}$ distribution using a generalized transition model with two characteristic mass scales, $\tilde{m}_{\rm low}$ and $\tilde{m}_{\rm high}$, allowing us to isolate spin populations within a finite interval of primary mass. The effective-spin distribution conditioned on primary mass is written as
\begin{eqnarray}
p(\chi_{\rm eff}\mid m_1)
=
p_{\rm out}(\chi_{\rm eff}\mid m_1)\,[1-\zeta(m_1)]
+\nonumber \\
p_{\rm in}(\chi_{\rm eff}\mid m_1)\,\zeta(m_1),
\end{eqnarray}
where $p_{\rm in}$ describes the spin population between the two transition masses and $p_{\rm out}$ describes the population outside this interval. The mixing function is defined as
\begin{equation}
\zeta(m_1)
= \small{\frac{1}{1+\exp\left[-(m_1-\tilde{m}_{\rm low})\over M_\odot\right]}\left[
1-\frac{1}{1+\exp\left[{-(m_1-\tilde{m}_{\rm high})\over M_\odot}\right]}\right].}
\end{equation}
In this construction, $\zeta(m_1)\simeq1$ for $\tilde{m}_{\rm low}\lesssim m_1\lesssim \tilde{m}_{\rm high}$ and $\zeta(m_1)\simeq0$ outside this interval. This generalized model therefore allows us to probe whether the effective-spin distribution changes within selected regions of the BBH mass spectrum.

The spin component $p_{\rm in}$ is modeled non-parametrically as

\begin{equation}
p_{\rm in}(\chi_{\rm eff}\mid m_1)
=
\frac{
\mathcal{H}(\chi_{\rm eff})
\,e^{\Theta(\chi_{\rm eff})}
}{
\int_{-1}^{1}
\mathcal{H}(\chi_{\rm eff})
\,e^{\Theta(\chi_{\rm eff})}
\,d\chi_{\rm eff}
},
\end{equation}
where $\Theta(\chi_{\rm eff})$ is modeled using a Gaussian process and $\mathcal{H}(\chi_{\rm eff})$ is a Heaviside window function defining the support of the distribution,

\begin{equation}
\mathcal{H}(\chi_{\rm eff})
=
\begin{cases}
1 & \chi_{{\rm min}}\leq \chi_{\rm eff}\leq \chi_{{\rm max}} \\
0 & \text{otherwise}
\end{cases}.
\end{equation}
This construction allows the minimum and maximum extent of each inferred $\chi_{\rm eff}$ distribution to vary freely and be constrained directly by the data. The parameters $\chi_{{\rm min}}$ and $\chi_{{\rm max}}$ are sampled independently for each population from the priors described in Table~\ref{tab:priors}. In particular, the lower bound is sampled conditionally on the upper bound according to

\begin{equation}
\pi(\chi_{{\rm min}}\mid\chi_{{\rm max}})
=
\chi_{{\rm min,unscaled}}
(\chi_{{\rm max}}+1)-1.
\end{equation}

The spin component outside the interval is modeled as a mixture of a Normal and a uniform distribution;
\begin{eqnarray}
p_{\rm out}(\chi_{\rm eff}\mid m_1)
=
\mathcal{N}(\chi_{\rm eff}; \, \mu, \, \sigma) \xi + \\
\mathcal{U}(\chi_{\rm eff}; \, \chi_{\rm min, out}, \, \chi_{\rm max, out})(1-\xi),
\end{eqnarray}
where $\mathcal{U}$ represents a uniform distribution with independent bounds, and
$\mathcal{N}$
is a truncated normal distribution.
The quantities $\mu$, $\sigma$, $\chi_{\rm min}$, and $\chi_{\rm max}$ are free parameters and sample according to Table~\ref{tab:priors}.
Since our primary interest is the population within the interval, these nuisance parameters are marginalized over but are not otherwise discussed or reported in this work. 

An important feature of this construction is that it allows the inferred distribution to approach arbitrarily small values over parts of parameter space if required by the data, i.e., very small (near zero) probabilities at certain points.
This provides more flexibility than commonly adopted parametric or spline-based models, which often enforce a non-zero density everywhere or impose a fixed functional form on the spin distribution. Our approach therefore enables the identification of localized features and disconnected support in the $\chi_{\rm eff}$ distribution that may arise from distinct astrophysical formation processes.

\section{\label{sec:Results} Results}

\begin{figure*}
\centering
\resizebox{1\textwidth}{!}{\includegraphics{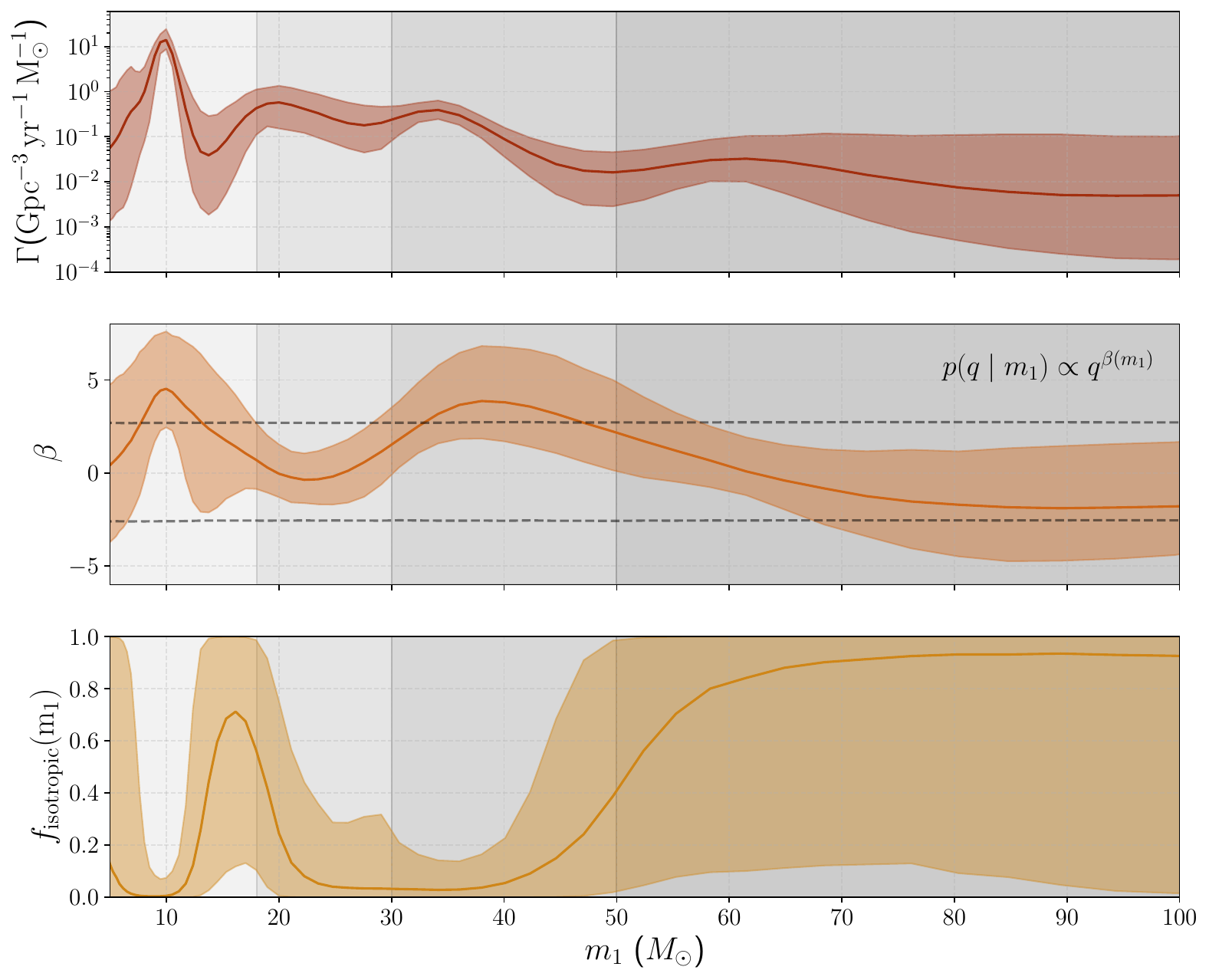}}
\caption{
Top panel: $\Gamma$ is  the astrophysical/source-frame merger-rate density per unit primary mass as a function of primary mass $m_1$. Middle panel: inferred slope $\beta(m_1)$ of the conditional mass-ratio distribution, $p(q\mid m_1)\propto q^{\beta(m_1)}$, obtained using our Gaussian-process population model. The dashed lines in this panel are the 90\% confidence bands of the prior on $\beta(m_1)$. Bottom panel: inferred mixing fraction of an isotropic spin population from the LVK analysis of Ref.~\citep{2026arXiv260527226T}, based on a model in which the $\chi_{\rm eff}$ distribution is described as a mixture between a Gaussian component and a broad uniform component, with the mixing fraction modeled as a Gaussian process as a function of mass. Vertical shaded bands indicate the mass intervals discussed in the text. Colored shaded regions show 90\% intervals, while the solid lines are the inferred median. Here we set
$(\tilde{m}_{\rm low},\,\tilde{m}_{\rm high})=(50,~200)M_\odot$. { However, the mass and $q$ distributions shown here were found to be independent of the choice for these parameters.}}
\label{fig:mergerrate_beta_iso}
\end{figure*}

\begin{figure}
\centering
\resizebox{0.49\textwidth}{!}{\includegraphics{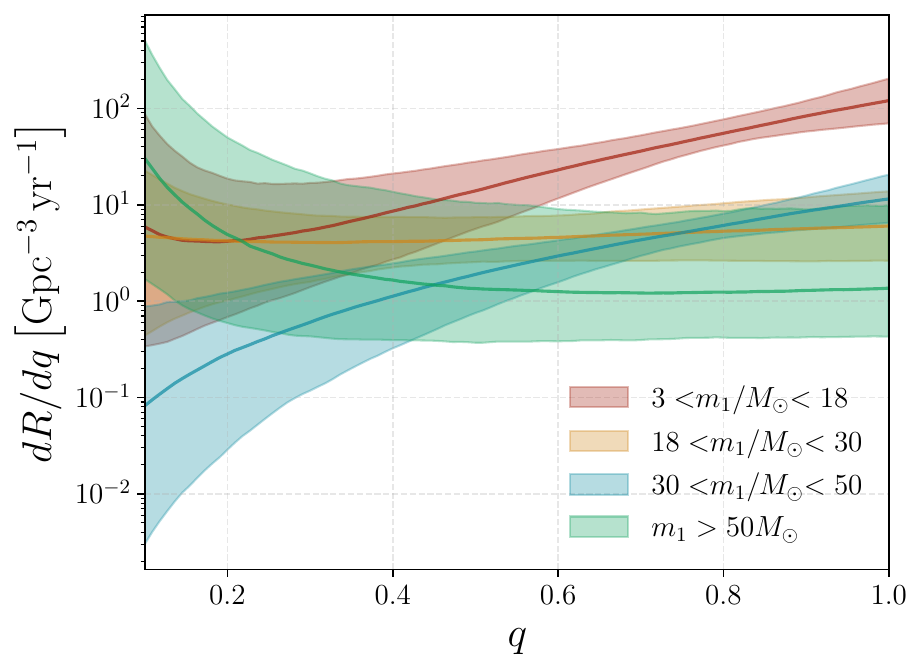}}
\caption{The merger rate as a function of mass ratio $q$ for each mass interval. The shaded region represents the 90\% confidence interval and the solid line is the median. Intervals $18<m_1/M_{\odot}<30$ and $m_1 > 50M_{\odot}$ show a flattening in the merger rate in comparison to $m_1 < 18M_{\odot}$ and $30<m_1/M_{\odot}<50$.}
\label{fig:qdist}
\end{figure}

\begin{figure*}
\centering
\resizebox{0.9\textwidth}{!}{\includegraphics{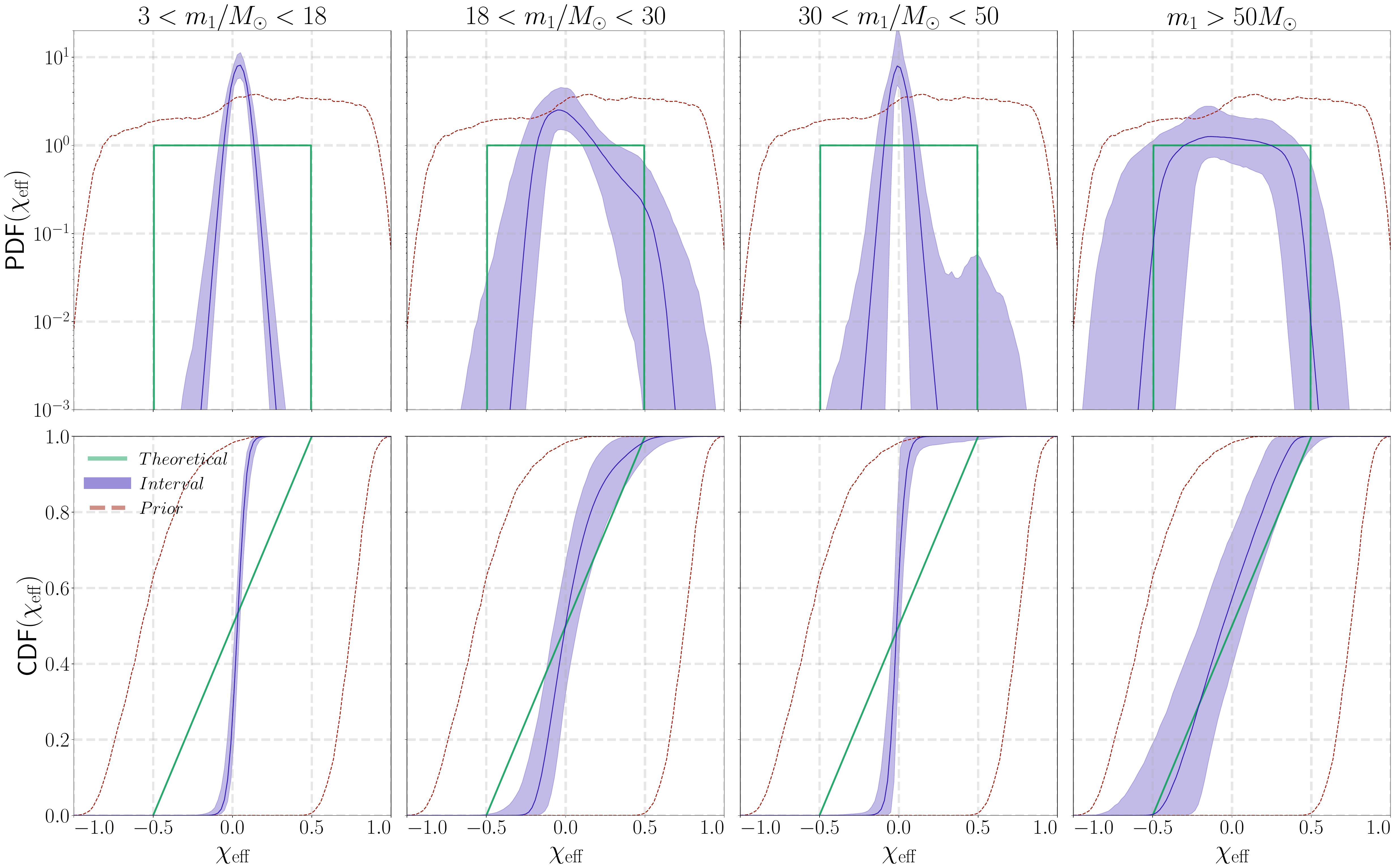}}
\caption{
Inferred effective-spin distributions across the four mass intervals identified from the mass-ratio analysis: $m_1<18\,M_{\odot}$, $18<m_1/M_{\odot}<30$, $30<m_1/M_{\odot}<50$, and $m_1>50\,M_{\odot}$. The top panels show the probability density functions and the bottom panels the corresponding cumulative distributions. 
Shaded regions show the $90\%$ confidence intervals, and solid lines the recovered median distribution. Red dotted lines show the prior ranges. The green lines indicate the distribution expected if the population formed hierarchically in dense clusters. 
The four intervals exhibit alternating narrow and broad spin populations, with the $m_1<18\,M_{\odot}$ and $30<m_1/M_{\odot}<50$ populations characterized by narrow distributions, while the $18<m_1/M_{\odot}<30$ and $m_1>50\,M_{\odot}$ intervals show substantially broader support extending toward both positive and negative effective spins.}
\label{fig:Xeff}
\end{figure*}

\subsection{Mass dependent mass-ratio distribution}

We first focus  on the dependence of the mass-ratio distribution as a function of primary mass. Figure~\ref{fig:mergerrate_beta_iso} shows the differential merger-rate density (top panel) together with the inferred slope $\beta(m_1)$ of the conditional mass-ratio distribution (middle panel).
The main result of this analysis is the emergence of four distinct mass regions characterized by  different mass-ratio distributions, which are shown in Figure~\ref{fig:qdist}.

At low masses, $m_1 \lesssim 18\,M_{\odot}$, the inferred slope satisfies $\beta \gtrsim 2$ at $90\%$ credibility, indicating a strong preference for nearly equal-mass binaries. This region coincides with the first peak in the merger-rate distribution around $10\,M_{\odot}$.

In the mass range $18$--$30\,M_{\odot}$, the inferred mass-ratio distribution becomes substantially flatter, with $\beta$ consistent with the interval $-1 \lesssim \beta \lesssim 1$. This behavior is inconsistent with the low-mass population at approximately the $90\%$ confidence level and coincides with a plateau-like feature in the merger-rate density.

At intermediate masses, within the interval $30$--$50\,M_{\odot}$, the population transitions back to a regime favoring equal-mass binaries, with $\beta \gtrsim 2$ once again preferred at $90\%$ confidence. This interval is associated with a possible secondary peak in the merger-rate distribution near $35\,M_{\odot}$ seen in the mass distribution.

Finally, above approximately $50\,M_{\odot}$, the inferred mass-ratio distribution flattens again, with $\beta$ returning to values consistent with $-1 \lesssim \beta \lesssim 1$. The inferred values of $\beta$ in this regime are statistically consistent with those found between $18$--$30\,M_{\odot}$, but inconsistent with the more equal-mass populations inferred below $\sim18\,M_{\odot}$ and between $30$--$50\,M_{\odot}$. This transition occurs near the sharp decline in the merger-rate density beginning at $\sim40\,M_{\odot}$, which is followed by a broad high-mass plateau.


 For comparison, the lower panel of Fig.~\ref{fig:mergerrate_beta_iso} shows the inferred mixing fraction of an isotropic spin population from the LIGO--Virgo--KAGRA collaboraton (LVK) analysis \citep{2026arXiv260527226T}, based on the model we introduced in \cite{antonini2026a}. In that framework, the $\chi_{\rm eff}$ distribution is described as a mixture between a Gaussian component and a broad uniform distribution extending over $|\chi_{\rm eff}|<0.5$, intended to represent an isotropic spin population such as might arise from hierarchical mergers. The mixing fraction between these two components is itself modeled as a Gaussian process as a function of primary mass.  

 
The transitions identified in $\beta(m_1)$ approximately coincide with mass intervals where the inferred isotropic spin fraction increases or decreases. In particular, in the flatter mass-ratio region above $50\,M_{\odot}$, the isotropic component is inferred to have a non-zero contribution at the $90\%$ confidence level. By contrast, in the intermediate mass ranges, where the population favors nearly equal-mass binaries, the isotropic fraction is constrained to remain at the level of $\lesssim10\%$.
At lower masses, the population below $\sim20\,M_{\odot}$ is largely dominated by a low-spin component with $f_{\rm isotropic}\simeq0$. However, toward the upper edge of this interval, a secondary isotropic component becomes visible, consistent with features previously identified in other studies \citep{antonini2026a, plunkett2026signaturessubpopulationhierarchicalmergers, tong2025subpopulationlowmassspinningblack}. This increase in the isotropic fraction coincides with a local dip in the merger-rate density, implying that this component contributes only a small fraction of the overall merger rate in this mass range.


Early LVK population analyses of GWTC-1 and GWTC-2 modeled the mass-ratio distribution using a single power law, $p(q)\propto q^\beta$, and generally inferred positive values of $\beta$, consistent with a preference for nearly equal-mass binaries \citep{Abbott_2021}. This is  consistent with our inference favoring large values of $\beta$ near $m_1\sim 10\,M_{\odot}$, where most of the astrophysical population is clustered. The GWTC-3 and GWTC-4  population analysis similarly found that the observed BBH population is dominated by comparable-mass systems, while also identifying confidently unequal-mass mergers such as GW190412 and GW190517\_055101 \citep{2023PhRvX..13a1048A, 2025arXiv250818083T}. However, these analyses largely assumed a smooth or weakly varying mass-ratio distribution across the full mass range.

The latest LVK GWTC-5 population analysis \citep{2026arXiv260527226T} has  identified evidence for more complex mass-dependent structure in the BBH population. In particular, that analysis reports that the feature near $35\,M_{\odot}$ is dominated by nearly equal-mass binaries, while the population above $45\,M_{\odot}$ exhibits substantially broader mass-ratio distributions. These results suggest that the shape of $p(q)$ evolves significantly across the primary-mass spectrum and that the BBH population may not be well described by a single smooth mass-ratio model.

This emerging picture is further supported by recent independent analyses. 
More recently, several studies have begun to identify evidence for multiple BBH subpopulations with distinct mass-ratio properties \citep{2026arXiv260317987R, 2025arXiv250915646B, 2025ApJ...993L..21A, Galaudage_2025}. \cite{2026arXiv260317987R}  found evidence for at least three BBH subpopulations separated by transitions in primary mass, including an intermediate-mass population strongly peaked toward equal masses and a high-mass population with broader mass-ratio distributions and larger spins. \cite{2025arXiv250915646B} and \cite{2026arXiv260525980Q} reached qualitatively similar conclusions,  arguing that the peak near $10\,M_{\odot}$ and the feature near $35\,M_{\odot}$ correspond to distinct populations with different mass-ratio and spin properties.

Our analysis supports the  results of previous and concurrent work although with some important differences. In particular, we recover evidence that the $35\,M_{\odot}$ feature is associated with a population strongly favoring equal-mass binaries, as seen in previous work \citep{sridhar2025characterizingbinaryblackhole, Li_2022, Li_2024, Kishore_Roy_2025, 2025arXiv250915646B, 2025arXiv250818083T, 2026arXiv260527226T, ray2026astrophysicaloriginbinaryblack, Farah:2026jlc, Vijaykumar:2026zjy}. Meanwhile, the population above $\sim45$--$50\,M_{\odot}$ exhibits substantially flatter mass-ratio distributions, consistent with the results of \cite{2025arXiv250915646B},  \cite{2026arXiv260317987R} and \cite{2026arXiv260525980Q}. However,  we also find that the low-mass BBH population itself separates into two statistically distinct regimes: a component below $18\,M_{\odot}$ characterized by strongly equal-mass binaries with $\beta\gtrsim2$, and a second component between $18$--$30\,M_{\odot}$ with substantially flatter mass-ratio distributions, $-1\lesssim\beta\lesssim1$. 
Our analysis therefore favors at least four distinct mass regimes separated by transitions in the mass-ratio distribution. These transitions coincide with  structures in the primary-mass spectrum, including the peak near $10\,M_{\odot}$, the plateau between $18$--$30\,M_{\odot}$, the feature near $35\,M_{\odot}$, and the high-mass tail above $50\,M_{\odot}$.

\subsection{Mass-dependent effective-spin distribution}

We consider  four population models in the following mass intervals: 
\[
(\tilde{m}_{\rm low},\,\tilde{m}_{\rm high})=
(3,18)\,M_{\odot}, 
\]
\[
(\tilde{m}_{\rm low},\,\tilde{m}_{\rm high})
=
(18,30)\,M_{\odot},
\]
\[
(\tilde{m}_{\rm low},\,\tilde{m}_{\rm high})
=
(30,50)\,M_{\odot},\rm 
\]
\[
(\tilde{m}_{\rm low},\,\tilde{m}_{\rm high})
=
(50,200)\,M_{\odot}~,
\]
representing the four regions identified from
the mass-ratio analysis in Figure~\ref{fig:mergerrate_beta_iso}.
Computing how many events contain at least 10\% of their primary mass posterior samples in each interval leaves 67 events in the $(3,18)\,M_{\odot}$ range, 85 in $(18,30)\,M_{\odot}$, 156 in $(30,50)\,M_{\odot}$, and 81 events for $(50,200)\,M_{\odot}$.
The inferred effective-spin distributions are given in 
Figure~\ref{fig:Xeff}. The inferred PDFs in adjacent  mass intervals show localized regions of strong posterior separation, with the density within and below/above the mass transition differing at $>99\%$ posterior probability at some values of $\chi_{\rm eff}$.
{ We stress here that we do not claim that the data uniquely select exactly three sharp change points at precisely the adopted boundary values. The chosen intervals provide a convenient discrete representation of the mass-dependent structure recovered in Figure~\ref{fig:mergerrate_beta_iso}. 
As a robustness check, in the Appendix we consider two alternative model choices and show that the inferred spin distributions are not sensitive to the precise transition masses adopted. In the first case, we retain two transitions per mass interval but shift their values. In the second
we use a single transition and vary its location across the mass spectrum. 
These additional analyses show that the main
 conclusions we derived are insensitive to the specific binning adopted.}

To quantify these differences, we compute the total-variation distance,
$
D_{\rm TV}
=
\frac{1}{2}
\int
\left|
p_i(\chi_{\rm eff})-p_j(\chi_{\rm eff})
\right|
\,{\rm d}\chi_{\rm eff}.
$
We find
$D_{\rm TV}=0.577^{+0.139}_{-0.238}$
between the $<18\,M_\odot$ and $[18,30]\,M_\odot$ intervals,
$D_{\rm TV}=0.556^{+0.258}_{-0.267}$
between the $[18,30]\,M_\odot$ and $[30,50]\,M_\odot$ intervals, and
$D_{\rm TV}=0.407^{+0.304}_{-0.221}$
between the $<18\,M_\odot$ and $[30,50]\,M_\odot$ intervals.
Both the integrated and pointwise comparisons therefore provide quantitative
evidence that the $\chi_{\rm eff}$ distribution in the
$[18,30]\,M_\odot$ interval differs from those in both adjacent mass regions.

The lowest-mass interval, $m_1\lesssim18\,M_{\odot}$, is characterized by a narrow $\chi_{\rm eff}$ distribution peaked at small positive values, with little evidence for extended tails. The distribution is approximately symmetric and closely resembles the population-averaged spin distribution inferred in previous LVK analyses, consistent with the fact that the majority of the BBH mergers in the astrophysical population are concentrated near the $10\,M_{\odot}$ peak and therefore dominate population-averaged measurements. In this mass  interval, 
the 5th percentile of the cumulative distribution function evaluated at zero lies above $0.15$, implying that more than $ 15\%$ of the population has negative effective spin at 95\% credibility.

In the second interval, $18\lesssim m_1/M_{\odot}\lesssim30$,
the inferred $\chi_{\rm eff}$ distribution changes significantly. The distribution becomes consistent with being centered at $\chi_{\rm eff}\simeq0$
and develops broader support toward both positive and negative values, together with a possible excess of positive $\chi_{\rm eff}\simeq 0.5$ systems or a positive skewness. 
Owing to the increased width of the distribution, however, the associated uncertainties grow significantly, and positive  values of the median comparable to those inferred for the $m_1\lesssim18~M_{\odot}$ population cannot be ruled out.
In the $18\lesssim m_1/M_{\odot}\lesssim30$ mass  interval, 
we find that more than $32\%$ of the population has negative effective spin at 95\% credibility.

The third interval, $30\lesssim m_1/M_{\odot}\lesssim50$, transitions back to a narrow $\chi_{\rm eff}$ distribution. In contrast to the first interval, however, the distribution is now sharply peaked around $\chi_{\rm eff}\simeq0$ rather than at positive values. The median and mean $90\%$ credible intervals do not overlap with those inferred for the lower-mass population,
providing strong evidence that the two populations are statistically distinct. 
Remarkably, the width of the distribution in this mass interval is nearly identical to that inferred for the $m_1\lesssim18\,M_{\odot}$ population. This may suggest that both populations are predominantly composed of first-generation black holes formed through similar stellar-evolution processes, while differing primarily in the mechanism through which binaries are assembled and merged. In particular, the higher-mass population may contain a larger contribution from dynamical assembly channels, naturally leading to more isotropic spin orientations despite retaining similarly narrow intrinsic spin magnitudes.
In the $30\lesssim m_1/M_{\odot}\lesssim50$ mass  interval, 
we find that more than $41\%$ of the population has negative effective spin at 95\% credibility.

Finally, the highest-mass interval, $m_1\gtrsim50\,M_{\odot}$, exhibits the broadest spin distribution, with support extending across a large fraction of the allowed $\chi_{\rm eff}$ range. The corresponding cumulative distribution becomes substantially flatter and consistent with expectations for an isotropic-spin population associated with hierarchical mergers.
In this highest mass interval we infer that more than $39\%$ of the population has negative effective spin at 95\% credibility.

\begin{figure}
\centering
\resizebox{0.5\textwidth}{!}{\includegraphics{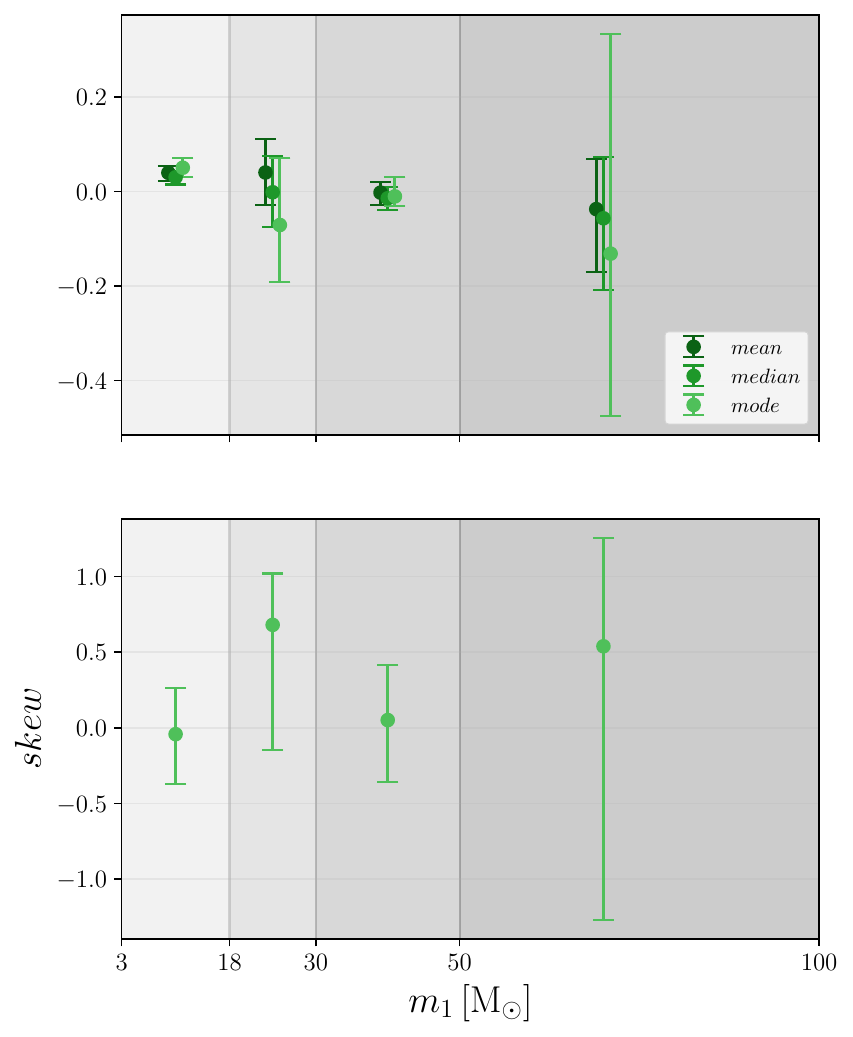}}
\caption{Summary statistics of the inferred effective-spin distributions across the four mass intervals identified from the mass-ratio analysis: $m_1<18\,M_{\odot}$, $18<m_1/M_{\odot}<30$, $30<m_1/M_{\odot}<50$, and $m_1>50\,M_{\odot}$. Error bars indicate the $90\%$ credible intervals. The upper panel shows the mean, median, and mode of the inferred $\chi_{\rm eff}$ distributions, while the lower panel shows the corresponding Pearson first skewness coefficient. The latter
is computed with respect to the mode of the distribution.
}
\label{fig:meanmedskew_regions}
\end{figure}

\begin{table*}
\centering
\begin{tabular*}{\textwidth}{@{\extracolsep{\fill}} l c c c c}
\hline
Interval & Mean & Median & Mode & Skew\\ 
\hline
$m_1<18\,M_{\odot}$ & $0.040_{-0.02}^{+0.01}$ & $0.030_{-0.02}^{+0.01}$ & $0.051_{-0.02}^{+0.02}$ & $-0.042_{-0.33}^{+0.31}$\\
$18<m_1/M_{\odot}<30$ & $0.040_{-0.07}^{+0.07}$ & $-0.001_{-0.07}^{+0.08}$ & $-0.071_{-0.12}^{+0.14}$ & $0.680_{-0.82}^{+0.34}$ \\
$30<m_1/M_{\odot}<50$ & $-0.002_{-0.03}^{+0.02}$ & $-0.014_{-0.02}^{+0.02}$ & $-0.010_{-0.02}^{+0.04}$ & $0.051_{-0.41}^{+0.36}$ \\
$m_1>50\,M_{\odot}$ & $-0.037_{-0.13}^{+0.11}$ & $-0.057_{-0.15}^{+0.13}$ & $-0.131_{-0.32}^{+0.46}$ & $0.539_{-1.81}^{+0.72}$ \\
\hline
\end{tabular*}
\caption{Results of summary statistics from the effective spin distribution for each mass interval. These are shown graphically in Figure~\ref{fig:meanmedskew_regions}.}
\label{tab:summarystats}
\end{table*}


\begin{figure}
\centering
\resizebox{0.48\textwidth}{!}{\includegraphics{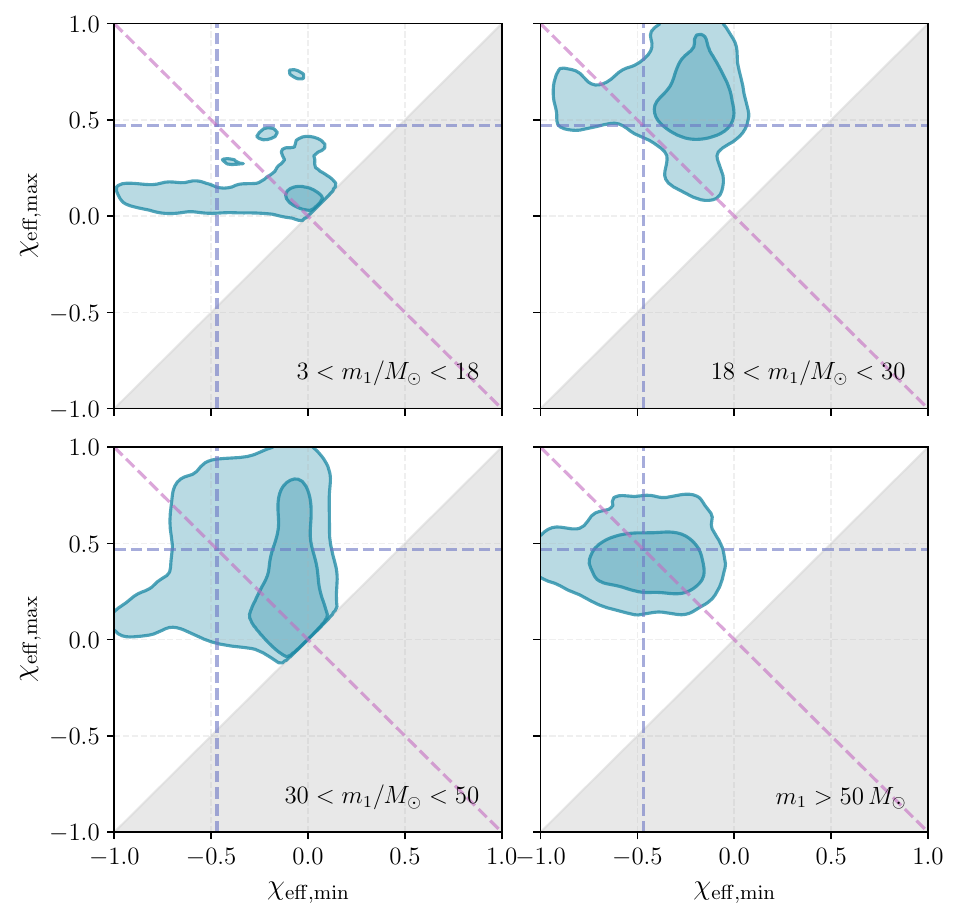}}
\caption{Posterior support for the minimum and maximum extent of the inferred effective-spin distributions across the four mass intervals identified from the mass-ratio analysis: $m_1<18\,M_{\odot}$, $18<m_1/M_{\odot}<30$, $30<m_1/M_{\odot}<50$, and $m_1>50\,M_{\odot}$. Contours at 95\% and 68\% confidence show the inferred joint posterior on $\chi_{\rm eff,min}$ and $\chi_{\rm eff,max}$ for each interval. The low-mass population is strongly concentrated toward small values of $\chi_{\rm eff,max}$, while the higher-mass intervals exhibit progressively broader support extending toward larger positive and negative effective spins. The blue dashed lines indicate $\chi_{\rm eff,min}=-0.47$ and $\chi_{\rm eff,max}=0.47$, approximately corresponding to the characteristic support expected from hierarchical mergers. 
Since for $m_1<18\,M_{\odot}$,  the expected value lies outside  the  population  
$90\%$ contours, in this mass range the data do not require a subdominant population of hierarchical mergers. A possible contribution is not excluded in any of the higher mass intervals. The pink dashed line marks the locus $y=-x$, corresponding to support intervals that are symmetric about $\chi_{\rm eff}=0$. They light gray region marks the forbidden region where $\chi_{\rm eff, min} > \chi_{\rm eff, max}$.
}
\label{fig:xeff_bounds}
\end{figure}

\subsubsection{Distribution symmetry}
Non-Gaussian features in the $\chi_{\rm eff}$ distribution like skewness, asymmetry about
zero, and multimodality can naturally arise in the $\chi_{\rm eff}$ distribution if multiple channels are contributing
to the detected population \citep{2025ApJ...990..147B}.
Thus, in this section we further investigate the symmetry of the recovered $\chi_{\rm eff}$ distributions.  In Figure~\ref{fig:meanmedskew_regions} we show  the mean, median, and mode of the distributions, while the lower panel shows the Pearson first skewness coefficient,
$\gamma_{\rm P}
=
\frac{\mu-\mathrm{Mode}}{\sigma},
$
where $\mu$ and $\sigma$ are the mean and standard deviation of the distribution. 
In Table~\ref{tab:summarystats} we report the recovered values for these quantities.


The lowest-mass population exhibits a small but consistently positive mean and median, confirming that the $\sim10\,M_{\odot}$ population is mildly shifted toward positive $\chi_{\rm eff}$. At the same time, the skewness remains consistent with zero, indicating that the distribution is symmetric around its peak. The broader $18$--$30\,M_{\odot}$ population shows tentative evidence for positive skewness, consistent with a possible excess near positive $\chi_{\rm eff}$, although a symmetric distribution remains consistent with the data.
The $30$--$50\,M_{\odot}$ interval is narrowly distributed and centered at $\chi_{\rm eff}\simeq0$, despite having a width comparable to that of the lowest mass population. 
 The highest-mass interval exhibits substantially larger uncertainties, primarily because the inferred distribution is much broader, making asymmetries difficult to constrain. Overall, all four mass intervals remain statistically consistent with symmetric $\chi_{\rm eff}$ distributions.

We note that previous work  identified a positive skewness in the overall $\chi_{\rm eff}$ distribution of the BH population \citep[e.g.,][]{2025ApJ...990..147B, 2026arXiv260527226T}. In the Appendix we show that this feature is most likely driven by binaries within primary masses  $\gtrsim 15 M_{\odot}$. Therefore our results suggest that, if present, the skewness is a mass dependent feature of the population.
 A
similar conclusion was reached in concurrent studies
\citep{alvarezlopez2026evidenceadditionalstructureeffective}. One possible interpretation is that
this feature reflects the emergence of a subpopulation of hierarchical
mergers, since remnants from previous mergers are expected to have
dimensionless spins of order $\sim 0.7$, naturally producing systems with
large spins and $\chi_{\rm eff}\lesssim 0.5$ when partially aligned with the
orbital angular momentum. However, we caution that such a feature is
intrinsically difficult to measure robustly. Spin measurements are subject to
large statistical uncertainties
\citep{vanDerSluys:2008qx,Vitale:2014mka,Vitale:2017cfs,Ghosh:2016qgn,
Chatziioannou:2018vzf,Pratten:2020ceb,Biscoveanu:2021wlv}, and the
mass-ratio--spin degeneracy can bias the inferred $\chi_{\rm eff}$ of
individual systems toward positive values
\citep{Cutler:1994ys,Poisson:1995ef,Baird:2012cu,Purrer:2013jcp,Ng:2018szy}.
In addition, selection effects preferentially enhance the detectability of
systems with large positive $\chi_{\rm eff}$ while disfavoring systems with
negative $\chi_{\rm eff}$
\citep{Flanagan:1997sx,Campanelli:2006uy,Ng:2018szy}. These effects can make
a broad underlying spin distribution appear skewed toward positive
$\chi_{\rm eff}$, particularly when the number of events contributing in a
given mass interval is small. Our hierarchical Bayesian analysis accounts for
selection effects at the population level, and  finds weak evidence  for positive skewness. Consequently, the apparent tail towards positive $\chi_{\rm eff}$ may reflect a broader isotropic  population whose negative-spin counterpart is more
difficult to resolve observationally.

\begin{figure*}
\centering
\resizebox{0.8\textwidth}{!}{\includegraphics{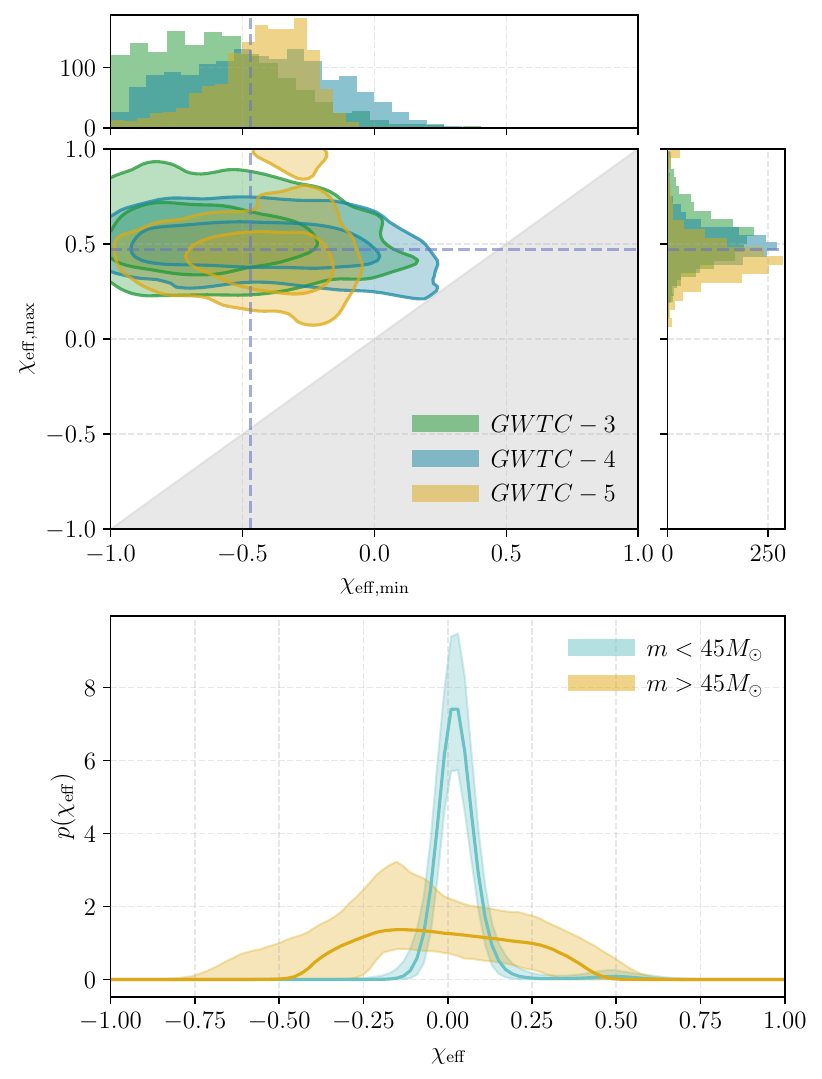}}
\caption{
Comparison between the inferred effective-spin support of the high-mass BBH population and the expectations for hierarchical mergers. The upper panel shows the joint posterior on $\chi_{\rm eff,min}$ and $\chi_{\rm eff,max}$ inferred from GWTC-3, GWTC-4, and GWTC-5 for the high-mass population above the transition mass scale $\tilde{m}=45M_\odot$. The corresponding marginalized distributions are shown along the top and right panels. With the increase of the catalog size, the constraints on the extent of the $\chi_{\rm eff}$ distribution have improve substantially. The data favor $\chi_{\rm eff,max}\sim0.5$ and increasingly constrain $\chi_{\rm eff,min}$ toward values near $-0.5$, while strongly disfavoring both $\chi_{\rm eff,min}=-1$ and $\chi_{\rm eff,min}=0$. The dashed lines indicate $\chi_{\rm eff,min}=-0.47$ and $\chi_{\rm eff,max}=0.47$, corresponding to the characteristic support expected from hierarchical mergers.
The lower panel compares the inferred $\chi_{\rm eff}$ distributions below and above the transition mass scale $\tilde{m}$. The low-mass population remains sharply peaked around small positive values of $\chi_{\rm eff}$, while the high-mass population develops substantially broader support extending toward both positive and negative effective spins. The broad support of the high-mass population is consistent with expectations for binaries containing remnants of previous black hole mergers. They light gray region marks the region where $\chi_{\rm eff, min} > \chi_{\rm eff, max}$.
}
\label{fig:mCut45}
\end{figure*}

\subsubsection{Constraints on  hierarchical merger populations}
Figure~\ref{fig:xeff_bounds} further illustrates the qualitative differences between the four mass intervals through the inferred support of the effective-spin distributions. The low-mass population, $m_1<18\,M_{\odot}$, is strongly concentrated toward small values of both $\chi_{\rm eff, min}$ and $\chi_{\rm eff,max}$, with little support for extended positive-spin tails. In particular, the data disfavor broad spin distributions extending toward $\chi_{\rm eff,max}\gtrsim0.5$, indicating that current observations do not require a significant contribution from hierarchical mergers to the dominant low-mass population. The absence of a broad high-spin component below $18\,M_{\odot}$ therefore suggests that the $10\,M_{\odot}$ peak is primarily composed of first-generation binaries formed through channels producing relatively small spins with a mild preference for aligned systems.

By contrast, the higher-mass intervals show progressively broader support extending toward larger values of $\chi_{\rm eff,max}$, especially for $18<m_1/M_{\odot}<30$ and $m_1>50\,M_{\odot}$. These regions are more compatible with populations containing hierarchical mergers or isotropic-spin systems. In particular, the data do not exclude the presence of binaries whose components are themselves remnants of previous black hole mergers.

Figure~\ref{fig:mCut45} illustrates how the inferred $\chi_{\rm eff}$ distribution for $m_1\gtrsim 45M_\odot$ compares with expectations for hierarchical mergers and how the observational constraints have improved from GWTC-3 to GWTC-5. 
This distribution was obtained as in the other models 
setting $
(\tilde{m}_{\rm low},\,\tilde{m}_{\rm high})
=
(45,200)\,M_{\odot}~$. We set  $\tilde{m}_{\rm low}=45M_\odot$ as this is a value close to the mass transition obtained in previous work
\citep{2025PhRvL.134a1401A,nxnr-pdyx,2025arXiv250210780T}, allowing for direct comparison with the literature.
The upper panel shows the inferred support of the minimum and maximum extent of the $\chi_{\rm eff}$ distribution obtained from successive gravitational-wave catalogs.

The upper bound of the distribution, $\chi_{\rm eff,max}$, is now  well constrained and remains consistent with the values expected from hierarchical mergers. At the same time, the lower bound $\chi_{\rm eff,min}$ is becoming increasingly constrained as the catalog grows, with the posterior now excluding  both $\chi_{\rm eff,min}=-1$ and $\chi_{\rm eff,min}=0$ at $99 \, \%$ confidence. This indicates that the observed high-mass  population is  fully consistent with a pure hierarchical formation scenario.

The lower panel directly compares the inferred $\chi_{\rm eff}$ distributions below and above the transition mass scale $\tilde{m}$. The low-mass population remains sharply peaked around small positive values of $\chi_{\rm eff}$ with very limited support for broad tails, while the high-mass population develops substantially broader support extending toward both positive and negative effective spins. Overall, these results are in qualitative agreement with our previous analysis, but the significantly larger GWTC-5 catalog now provides substantially tighter constraints on the allowed extent and asymmetry of the high-mass spin population.

\section{\label{sec:Discussion} Astrophyiscal Interpretation and Conclusions}


In this section, we discuss the astrophysical implications of the mass-dependent structure identified in the BBH population. Our analysis reveals multiple  transitions in both the mass-ratio and effective-spin distributions as a function of primary mass, suggesting that different regions of the BBH mass spectrum are governed by different binary properties and possibly by different formation scenarios.  We consider here whether the transitions identified in $p(q|m_1)$ and $p(\chi_{\rm eff}|m_1)$ reflect changes in the dominant mechanisms responsible for binary assembly and evolution.

A particularly important result of this work is the strong correlation between changes in the mass-ratio distribution and corresponding changes in the effective-spin distribution. The same mass scales that separate equal-mass and broad mass-ratio populations also coincide with transitions from narrow aligned-spin distributions to broader distributions consistent with isotropic or hierarchical populations. This connection provides further evidence that the evolution of binary mass ratios and spins is closely linked and that both quantities carry complementary information about the astrophysical origin of BBH mergers.

A clear transition occurs near $18~M_{\odot}$, where the effective-spin distribution shifts from a narrow distribution peaked at positive $\chi_{\rm eff}$ to a broader distribution centered near zero. 
Here we fix the transition mass scale based on the independent change observed in $p(q|m_1)$; in a companion study \citep{flanagan+2}, we show that the mass value of the  transition  can be inferred directly from the data through a change in $p(\chi_{\rm eff}|m_1)$ alone.
Together, these results indicate that the population associated with the $10~M_{\odot}$ peak is statistically distinct from the higher mass population.

The low-mass population is characterized by relatively small spins with a mild preference for positive $\chi_{\rm eff}$, while still retaining significant support at negative values. Such a distribution might be achieved  with standard isolated binary evolution models. 
In isolated binary evolution, tidal synchronization, mass transfer, and common-envelope interactions tend to align stellar spins with the orbital angular momentum prior to collapse, naturally producing BBHs with preferentially positive $\chi_{\rm eff}$~\citep{Qin_2018, Bavera2020, Fuller2019, Belczynski2020}.
At the same time, binary stellar evolution naturally favors nearly equal-mass systems because mass transfer and common-envelope evolution efficiently drive binaries toward comparable component masses~\citep{Belczynski2008,Dominik2012,deMink2015,Marchant2016}. Natal kicks during compact-object formation or/and inefficient tides can nevertheless result in significant  spin-orbit misalignment, broadening the $\chi_{\rm eff}$ distribution 
and producing some finite support for negative values \citep{Qin_2018,Bavera2020}. Overall, however, matching the narrow mildly positive $\chi_{\rm eff}$ distribution of the low-mass population within standard isolated binary evolution models might require  non-standard assumptions~\citep{flanagan+2} given the  significant fraction of misaligned systems required by the data.

Similarly, triple-mediated formation channels are expected to  produce nearly equal-mass binaries since the stellar processes that drive isolated binaries also affect the evolution of the inner binary in a triple system. 
Although previous work has shown that some fraction of the population may exhibit a broader mass-ratio distribution \citep{2022PhRvD.106b3014S}, we note that those models neglected binary interactions, which are the main mechanism expected to drive the preference for equal-mass binaries in this scenario \citep{2022PhRvD.106b3014S}.
 Secular dynamical interactions in triples can significantly misalign the orbital plane and black-hole spins, naturally producing a population with a mild bias toward positive $\chi_{\rm eff}$
 while still retaining substantial support for anti-aligned systems \citep{2018MNRAS.480L..58A,2026ApJ..1000L..59S}. Therefore, triple mediated mergers might also explain the observed distribution \citep{flanagan+2}.

The intermediate region between
 $18$ and $30\,M_\odot$
 is more difficult to interpret. Here, the population exhibits broader mass-ratio and spin distributions together with tentative evidence for positive skewness and an excess near $\chi_{\rm eff}\sim0.5$. Importantly, the inferred positive skewness does not necessarily imply that the population is dominated by positively aligned systems. Instead, the skewness appears to be driven by a possible extended tail toward $\chi_{\rm eff}\sim0.5$, while the bulk of the distribution remains concentrated near zero spin. Such a feature may naturally arise from a broader isotropic component associated with hierarchical mergers, whose negative-spin counterpart is more difficult to detect because binaries with positive $\chi_{\rm eff}$ remain longer in band and are observationally favored. 
 Consequently, although skewness is possible in this region, the current data remain statistically consistent with a symmetric effective spin distribution (see Fig.~\ref{fig:meanmedskew_regions}).

One possibility is therefore that the $18$--$30~M_{\odot}$ interval represents a transition between the low-mass mildly aligned population and the dynamically assembled populations dominating at higher masses. Alternatively, this region may indicate the emergence of an additional channel, such as mergers embedded in AGN disks~\citep[e.g.,][]{2021ApJ...908..194T, Wang2021ApJ...923L..23W,alvarezlopez2026evidenceadditionalstructureeffective} or the onset of hierarchical mergers below the pair-instability scale \citep[e.g.,][]{2023MNRAS.520.5259A, 2025ApJ...993L..21A}. In the former scenario, the excess of positive $\chi_{\rm eff}$ systems could arise from accretion torques within the AGN disk partially aligning the spins of compact-object binaries with the orbital angular momentum. However, current AGN models do not naturally explain why the distribution would terminate near $\chi_{\rm eff}\sim0.5$. By contrast, in the hierarchical-merger interpretation, this characteristic scale arises naturally from the spins expected for merger remnants. In this case, the apparent lack of a corresponding negative-$\chi_{\rm eff}$ tail may simply reflect the observational bias against detecting and characterizing anti-aligned systems as discussed above.

At higher masses, near the feature around $m_1 \simeq 35~M_{\odot}$, the width of the
$\chi_{\rm eff}$ distribution is nearly identical to that inferred for the low-mass population below $18M_\odot,  $ despite the shift in the location of the peak at zero. This may indicate that both regions are dominated by first-generation black holes formed through similar stellar-evolution processes, while differing primarily in the mechanism through which binaries are assembled. In particular, the $35~M_{\odot}$ population may naturally arise from dynamical pairing of first-generation black holes in dense stellar clusters \citep{2023MNRAS.522..466A, 2022MNRAS.513.4527C}, producing binaries with small intrinsic spins but nearly isotropic spin-orbit orientations and, possibly, an extended tail representing a subdominant population of hierarchical mergers.

Above $m_1\simeq50~M_{\odot}$, both the mass-ratio and effective-spin distributions broaden substantially. This transition occurs close to the expected onset of the pair-instability mass gap reported in previous studies \citep{2017MNRAS.470.4739S,2019ApJ...887...53F,2019A&A...624A..66R,2023MNRAS.526.4130H} and coincides with the emergence of spin distributions fully consistent with hierarchical mergers. In this regime, repeated mergers in dense stellar clusters may dominate the observed population, naturally producing broader mass-ratio distributions together with effective-spin support extending toward both positive and negative values \citep{2023MNRAS.526.4908C}. The inferred support of the spin distribution is  consistent with binaries composed of first- and second-generation black holes \citep{2025PhRvL.134a1401A,nxnr-pdyx}. The skewness, mean, median and mode of the $\chi_{\rm eff}$ distribution are all consistent with being zero and therefore with an isotropic distribution of spin orientations as expected for dynamical formation in clusters \citep{Rodriguez2016c}.

The breadth of the high-mass effective-spin distribution matches the signature expected from repeated mergers. The remnant of a near-equal-mass merger acquires a nearly universal dimensionless spin peaked at $a\simeq0.7$, with little support below $a\simeq0.5$ \citep{2017ApJ...840L..24F}. For an isotropic distribution of spin orientations, this bounds the effective spin at $|\chi_{\rm eff}|\lesssim a/1.5\approx0.47$ and produces a broad, approximately symmetric distribution \citep{2025PhRvL.134a1401A}, in agreement with the support we infer above $50~M_{\odot}$. The coincidence of this transition with the inferred lower edge of the pair-instability mass gap, $\sim45~M_{\odot}$ \citep{antonini2026a, tong2026a}, strengthens the interpretation that the highest-mass systems are predominantly remnants of earlier mergers repopulating the gap. 

The fraction of the high-mass population produced hierarchically is governed by the escape velocity of the host environment, which determines whether merger remnants are retained against gravitational-wave recoil \citep{2023MNRAS.526.4908C}. Globular-cluster-like potentials ($v_{\rm esc}\sim50~{\rm km\,s^{-1}}$) retain only a small fraction of remnants, whereas nuclear star clusters ($v_{\rm esc}\sim250~{\rm km\,s^{-1}}$) retain a substantially larger fraction \citep{2021ApJ...918L..31M}. The relative scarcity of mergers above $50~M_{\odot}$ therefore limits the contribution of environments with the highest escape velocities, which would otherwise overproduce massive mergers \citep{2022arXiv220508549Z}.

We note that while the highest-mass spin distribution is naturally explained within the hierarchical-merger scenario, current data do not exclude distributions that exhibit either a positive or a negative excess of $\chi_{\rm eff}$ systems. Alternative formation channels, including stellar evolution with fallback-induced spin-up of the compact remnant \citep{Janka2013, Chan_2020}, stellar mergers \citep{2008MNRAS.383L...5G, 2021MNRAS.507.5132D}, chemically homogeneous evolution \citep{DallAmico2025, Riley2021, Mandel2016a, DeMink2016}, and formation within AGN \citep{Bartos2016,2021ApJ...908..194T}
accretion disks, all provide possible explanations for this population. However, unlike the hierarchical-merger hypothesis, these scenarios do not currently make robust or quantitative predictions for the shape and support of the $\chi_{\rm eff}$ distribution, making them considerably more difficult to test directly against gravitational-wave observations.

\bigskip

We thank Charlie Hoy for serving as our internal LVK reviewer. FA and FD are supported by the UK’s Science and Technology Facilities Council grants
ST/V005618/1 and UKRI2489. IMRS acknowledges support from the Science and Technology Facilities Council grant number ST/Y001990/1 and the Science and Technology Facilities Council Ernest Rutherford Fellowship grant number UKRI2423. DC acknowledges support from the Gordon and Betty Moore Foundation (Grant GBMF12341). 
This material is based upon work supported by
NSF’s LIGO Laboratory which is a major facility fully funded by
the National Science Foundation, as well as the Science and Technology Facilities Council (STFC) of the United Kingdom, the Max-Planck-Society (MPS), and the State of Niedersachsen/Germany for
support of the construction of Advanced LIGO and construction and
operation of the GEO600 detector. Additional support for Advanced
LIGO was provided by the Australian Research Council. Virgo is
funded, through the European Gravitational Observatory (EGO), by
the French Centre National de Recherche Scientifique (CNRS), the
Italian Istituto Nazionale di Fisica Nucleare (INFN) and the Dutch
Nikhef, with contributions by institutions from Belgium, Germany,
Greece, Hungary, Ireland, Japan, Monaco, Poland, Portugal, Spain.
KAGRA is supported by Ministry of Education, Culture, Sports, Science and Technology (MEXT), Japan Society for the Promotion of
Science (JSPS) in Japan; National Research Foundation (NRF) and
Ministry of Science and ICT (MSIT) in Korea; Academia Sinica
(AS) and National Science and Technology Council (NSTC) in Taiwan. This research has made use of data or software obtained from the Gravitational Wave Open Science Center (gwosc.org), a service of the LIGO Scientific Collaboration, the Virgo Collaboration, and KAGRA. The authors are grateful for computational resources provided
by Cardiff University and supported by STFC grant
ST/V005618/1.



\bibliography{ref}{}
\bibliographystyle{aasjournalv7}

\clearpage
\onecolumngrid

\setcounter{equation}{0}
\setcounter{figure}{0}
\setcounter{table}{0}
\setcounter{section}{0}

\renewcommand{\theequation}{S\arabic{equation}}
\renewcommand{\thefigure}{S\arabic{figure}}
\renewcommand{\thetable}{S\arabic{table}}

\renewcommand{\thesection}{}
\renewcommand{\thesubsection}{}

\begin{center}
{\Large \bf Appendix}
\end{center}

\vspace{1cm}

\begin{figure}
\centering
\resizebox{1.\textwidth}{!}{\includegraphics{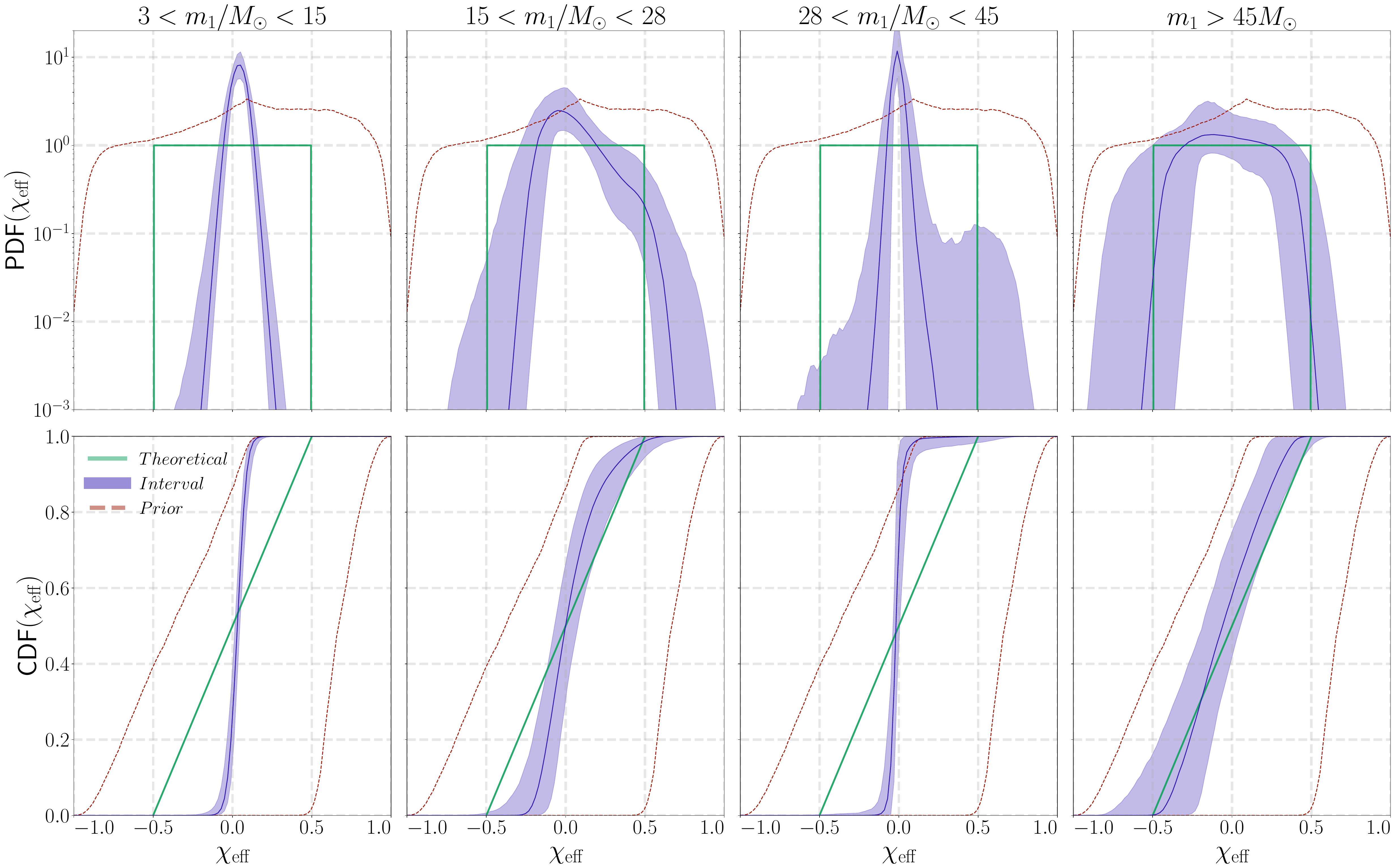}}
\caption{
 The mass transitions of each region has been shifted from the original choices. We see no significant change in comparing these to the originally selected mass cuts.
}
\label{fig:mCut_shifted}
\end{figure}

\begin{figure}
\centering
\resizebox{0.9\textwidth}{!}{\includegraphics{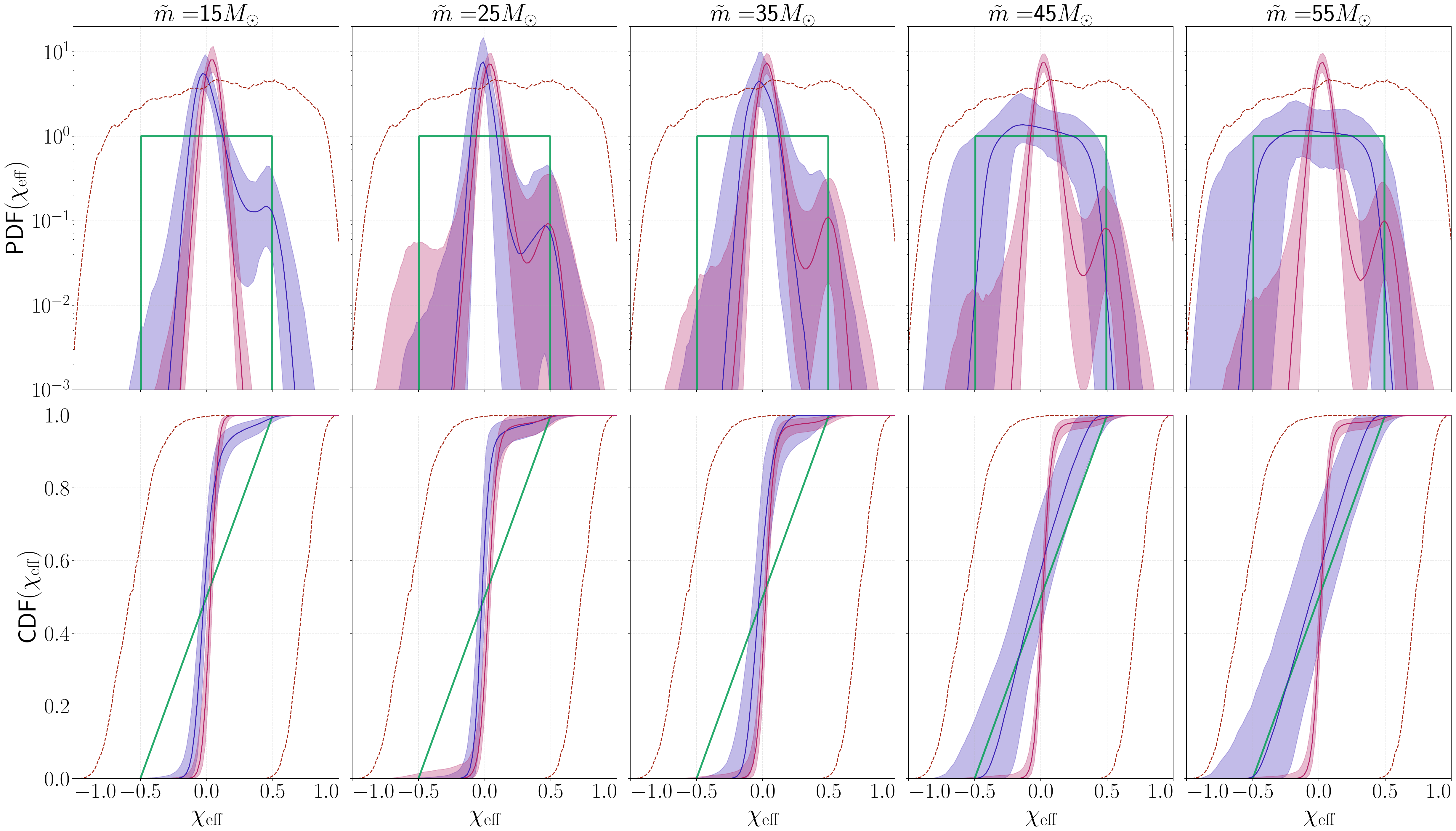}}
\caption{ 
Inferred effective-spin distributions  for different choices of the transition mass scale $\tilde{m}$. Columns correspond to $\tilde{m}=15,\,25,\,35,\,45,$ and $55\,M_{\odot}$, respectively. The top panels show the inferred probability density functions, while the bottom panels show the corresponding cumulative distribution functions. Blue and pink curves indicate the two populations below and above $\tilde{m}$, respectively.
Shaded regions denote the $90\%$ credible intervals, and solid lines the median.
The green curve shows the uniform distribution expected for hierarchical mergers in dense clusters \citep{2025PhRvL.134a1401A}.  The red dashed curves indicate the broad isotropic-spin priors used in the analysis.
}
\label{fig:Xeff_bigplot}
\end{figure}

\begin{figure}
\centering
\resizebox{0.5\textwidth}{!}{\includegraphics{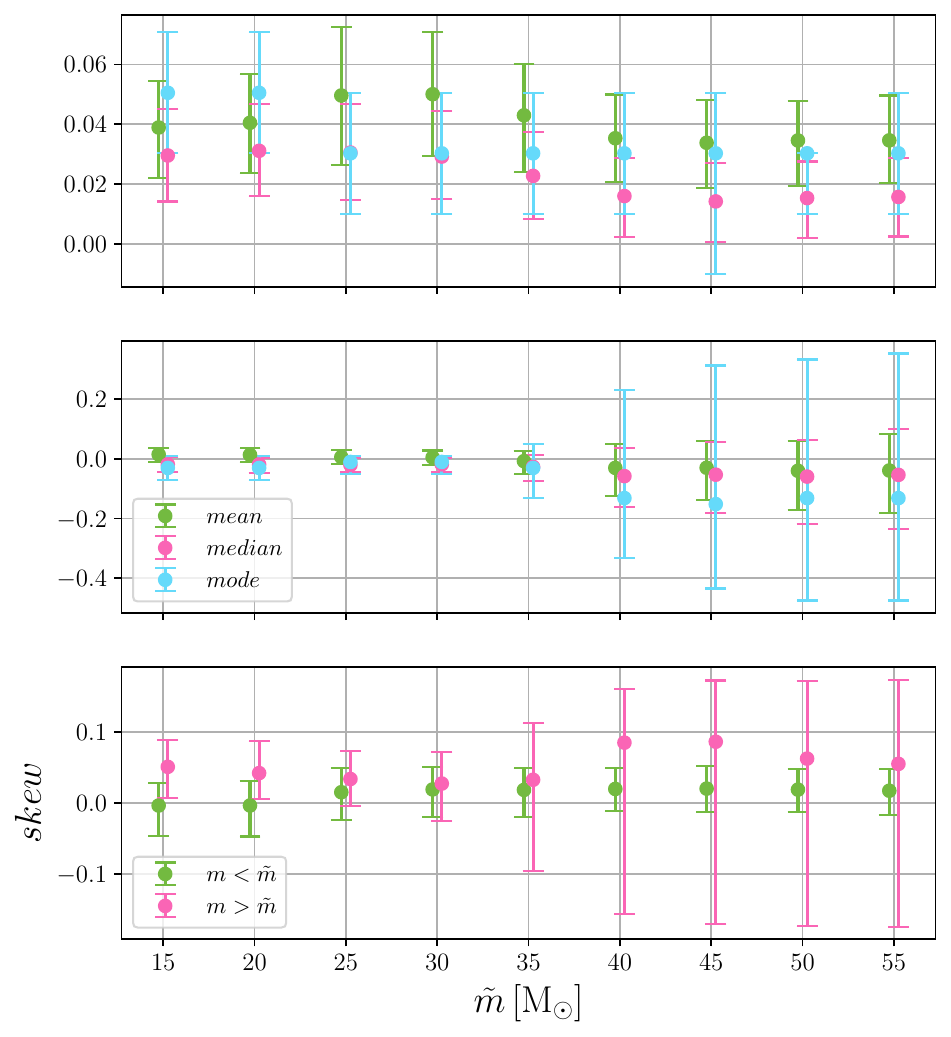}}
\caption{
The evolution of the inferred effective-spin distributions as a function of the transition mass scale $\tilde{m}$. Error bars indicate the $90\%$ credible intervals.
The top panel shows the inferred mean, median and mode of the effective spin of the  population with $m_1<\tilde{m}$. The middle panel shows the inferred mean, median, and mode of  the  population with $m_1>\tilde{m}$, illustrating the progressive shift of the high-mass population toward $\chi_{\rm eff}\simeq0$ and the increasing uncertainty associated with broader spin distributions at large masses. The bottom panel shows the Pearson first skewness coefficient, quantifying the degree of asymmetry of the inferred $\chi_{\rm eff}$ distributions. While all measurements remain statistically consistent with symmetric distributions, the intermediate-mass regime exhibits tentative evidence for positive skewness associated with the possible excess near $\chi_{\rm eff}\sim0.5$.
}
\label{fig:meanmedskew}
\end{figure}

\section*{Additional Spin Population Models}
{ Here we consider two additional sets of population models to assess the sensitivity of the inferred $\chi_{\rm eff}$ distributions to the adopted mass intervals. As a first test we repeat the analysis using four fixed mass intervals that differ from those used in the main text:
\[
3 \leq m_1/M_{\odot} < 15,
\]
\[
15 \leq m_1/M_{\odot} < 28,
\]
\[
28 \leq m_1/M_{\odot} < 45,
\]
\[
45 \leq m_1/M_{\odot} < 200.
\]
The $\chi_{\rm eff}$ distributions inferred in these alternative intervals are shown in Fig.~\ref{fig:mCut_shifted} are essentially unchanged from those shown in Fig.~\ref{fig:Xeff}.  This demonstrates that the qualitative spin structure identified in the main analysis is not sensitive to small changes in the adopted mass boundaries.}

In one further set of models, we describe the effective-spin distribution as a mixture of two independent Gaussian-process components separated by a single transition mass scale $\tilde{m}$. The first component describes binaries with primary masses below $\tilde{m}$, while the second describes binaries above $\tilde{m}$. Rather than inferring $\tilde{m}$ directly from the data, we fix its value and repeat the analysis for different choices of $\tilde{m}$. This approach  complements the approach presented in the main text and allows us to probe how the inferred $\chi_{\rm eff}$ distribution changes as progressively higher-mass binaries are included in the high-mass population, thereby identifying the mass scales at which transitions in the spin distribution occur. These results are consistent and reinforce the conclusions drawn in the main text.

Figure~\ref{fig:Xeff_bigplot} shows the inferred effective-spin distributions for different choices of the transition mass scale $\tilde{m}$. The blue and red curves correspond to the low- and high-mass subpopulations, respectively. By varying $\tilde{m}$, we directly probe how the inferred $\chi_{\rm eff}$ distribution changes with primary mass.

Below $15\,M_{\odot}$, the effective-spin distribution is well described by a narrow approximately Normal distribution with a small positive mean and little evidence for extended tails. In this regime, the data do not require a broad isotropic-spin component. This behavior is broadly consistent with previous population analyses of the entire population, likely because the majority of observed BBH mergers are concentrated within the $10\,M_{\odot}$ peak and therefore dominate population-averaged measurements.

Above $15\,M_{\odot}$, however, the inferred $\chi_{\rm eff}$ distribution changes significantly. The distribution becomes centered closer to $\chi_{\rm eff}\simeq0$ and develops broader support toward both positive and negative values, together with a possible excess of positive-$\chi_{\rm eff}\simeq 0.5$ systems. The data confidently exclude models in which the low- and high-mass spin populations are statistically consistent with one another.

For $\tilde{m}\gtrsim20\,M_{\odot}$, both the low- and high-mass populations develop a possible asymmetric tail with excess support near $\chi_{\rm eff}\sim0.5$, similar to features identified in recent concurrent studies \citep{2025ApJ...990..147B, 2025arXiv251018867R, plunkett2026signaturessubpopulationhierarchicalmergers,alvarezlopez2026evidenceadditionalstructureeffective}. Since this excess disappears for $\tilde{m}\lesssim15\,M_{\odot}$ and for $\tilde{m}\gtrsim30\,M_{\odot}$, it appears to be associated primarily with systems in the mass range $15\lesssim m_1/M_{\odot}\lesssim30$, in agreement with our main analysis.
Interestingly, $\chi_{\rm eff}\sim0.5$ is close to the value naturally expected from hierarchical mergers. When partially aligned with the orbital angular momentum, such systems can produce effective spins near the observed value. 


For transition masses $\tilde{m}\gtrsim45\,M_{\odot}$, the inferred high-mass population becomes substantially broader and develops support extending over a large fraction of the allowed $\chi_{\rm eff}$ range. The corresponding cumulative distributions flatten significantly, approaching the behavior expected for a broad isotropic-spin population associated with hierarchical mergers. By contrast, the low-mass population remains dominated by a narrow distribution centered near $\chi_{\rm eff}\simeq0.05$, with evidence for a broader possibly asymmetric component. The inferred high-mass spin distributions therefore become qualitatively consistent with the isotropic-spin population inferred in our previous work \citep{2025PhRvL.134a1401A,antonini2026a}.

Figure~\ref{fig:meanmedskew} provides a compact summary of how the inferred effective-spin distributions evolve as the transition mass scale $\tilde{m}$ is varied. The mean, median, and mode of the low mass population remain consistently positive and peaked around $\chi_{\rm eff}\simeq 0.03$, while the inferred skewness with respect to its mode stays consistent with zero. This behavior reflects the fact that the majority of BBH mergers are concentrated within the $10~M_{\odot}$ peak, which therefore dominates the inferred low-mass distribution whenever included in the analysis.

By contrast, the properties of the high-mass population evolve significantly as $\tilde{m}$ increases. For $\tilde{m}\lesssim30~M_{\odot}$, the high-mass population retains a small positive shift in $\chi_{\rm eff}$, but above this scale the distribution progressively moves toward $\chi_{\rm eff}\simeq0$ and becomes substantially broader. The increasing uncertainties in the mode and skewness reflect the emergence of broad spin distributions with extended support toward both positive and negative effective spins. While the inferred skewness remains statistically consistent with zero at all masses, the data show tentative evidence for a positive tail in the intermediate-mass regime, consistent with the possible excess towards $\chi_{\rm eff}\sim0.5$ identified in the full posterior distributions.
However, as discussed in the main text, this apparent excess may instead reflect the difficulty of accurately modeling selection effects that preferentially favor positive $\chi_{\rm eff}$ in sparsely populated regions of parameter space.


\section*{Hierarchical inference and data}


We perform hierarchical population inference using a Hamiltonian Monte Carlo (HMC) method, implemented in \texttt{numpyro} \citep{phan2019composable} using \texttt{jax} \citep{jax2018github}, following the procedure laid out in \citep{nxnr-pdyx}. We begin by defining the posterior probabilities for hyperparameters $\Lambda$, given a set of observed data $\{d_{i}\}$

\begin{equation}
    p(\Lambda | \{d_{i}\}) \propto 
    p(\Lambda) \,
    \xi^{-N_{\text{det}}}(\Lambda)\prod^{N_{\text{det}}}_{i=1}
    \biggr \langle \cfrac{p(\theta_{i} \mid \lambda)}{p_{\text{pe}}(\theta_{i})} \biggr \rangle .
    \label{eq:pop_likelihood}
\end{equation}

Where $p(\Lambda)$ are the population parameter priors and $p_{\text{pe}}(\theta_{i})$ is the single-event parameter estimation prior. We take an average over the posterior samples for each event $i$ \citep{nxnr-pdyx, Fishbach_2018}. Selection effects are accounted for using the detection efficiency;

\begin{equation}
    \xi(\Lambda) = \cfrac{1}{N_{\text{inj}}} \sum_{i=1}^{N_{\text{found}}} \cfrac{p(\theta_{i}|\Lambda)}{p_{\text{inj}}(\theta_{i})}.
\end{equation}

This re-weights the recovered injections from a reference injection distribution $p_{\text{inj}}(\theta_{i})$, giving the total fraction of events expected to pass our detection criteria.

We track the number of effective posterior samples $N_{\rm eff}$ for each event to control Monte Carlo noise while sampling

\begin{equation}
    N_{\text{eff}}(\Lambda) = \cfrac{\left[\sum_{j}w_{i,j}(\Lambda)\right]^{2}}{\sum_{j}w_{i,j}^{2}(\Lambda)},
    \hspace{1cm}
    w_{i, j} = \cfrac{p(\theta_{i,j} | \Lambda)}{p_{\text{pe}}(\theta_{i,j})},
\end{equation}

and similarly for the injections we track,

\begin{equation}
    N^{\text{inj}}_{\text{eff}}(\Lambda) = \cfrac{\left[\sum_{i}w_{i}(\Lambda)\right]^{2}}{\sum_{i}w_{i}^{2}(\Lambda)}.
\end{equation}

Following \citep{essick2022precisionrequirementsmontecarlo}, we require $N^{\text{inj}}_{\text{eff}}(\Lambda) \gtrsim 4N_{\text{det}}$. As opposed to hard-cuts, we add penalties to the log-likelihood to exclude poorly sampled regions of the parameter space,

\begin{equation}
    \text{ln}\,S\left(\cfrac{N^{\text{inj}}_{\text{eff}}(\Lambda)}{4N_{\text{det}}}\right) + \text{ln}\,S\left(\cfrac{\mathcal{N}}{0.6}\right),
    \hspace{1cm}
    S(x) = \cfrac{1}{1 + x^{-30}},
    \hspace{1cm}
    \mathcal{N} \equiv \text{min}_{i} \, \text{log} N_{\text{eff},i}.
\end{equation}

This allows for the inference to smoothly suppress models where either the injection set or the events are represented by too few effective samples.

For events from GWTC-1 \citep{2019PhRvX...9c1040A}, we use the parameter estimation samples \textit{Overall posterior}. For sources found in GWTC-2 \citep{Abbott_2021}, we use the parameter estimation samples \textit{PrecessingSpinIMRHM}, and for events found in GWTC-3 \citep{PhysRevX.13.041039}, we use \textit{C01:Mixed} samples. For GWTC-4 \citep{LVKcat_inprep} and GWTC-5 \citep{2026arXiv260527225T} events, we preferentially use \texttt{NRSur7dq4}. In the case of those samples being unavailable, we use \textit{Mixed} samples for GWTC-4 and \textit{IMRPhenomXPHM-SpinTaylor} \cite{PhysRevD.103.104056, PhysRevD.111.104019} samples for GWTC-5.

\begin{figure}
\centering
\resizebox{1.\textwidth}{!}{\includegraphics{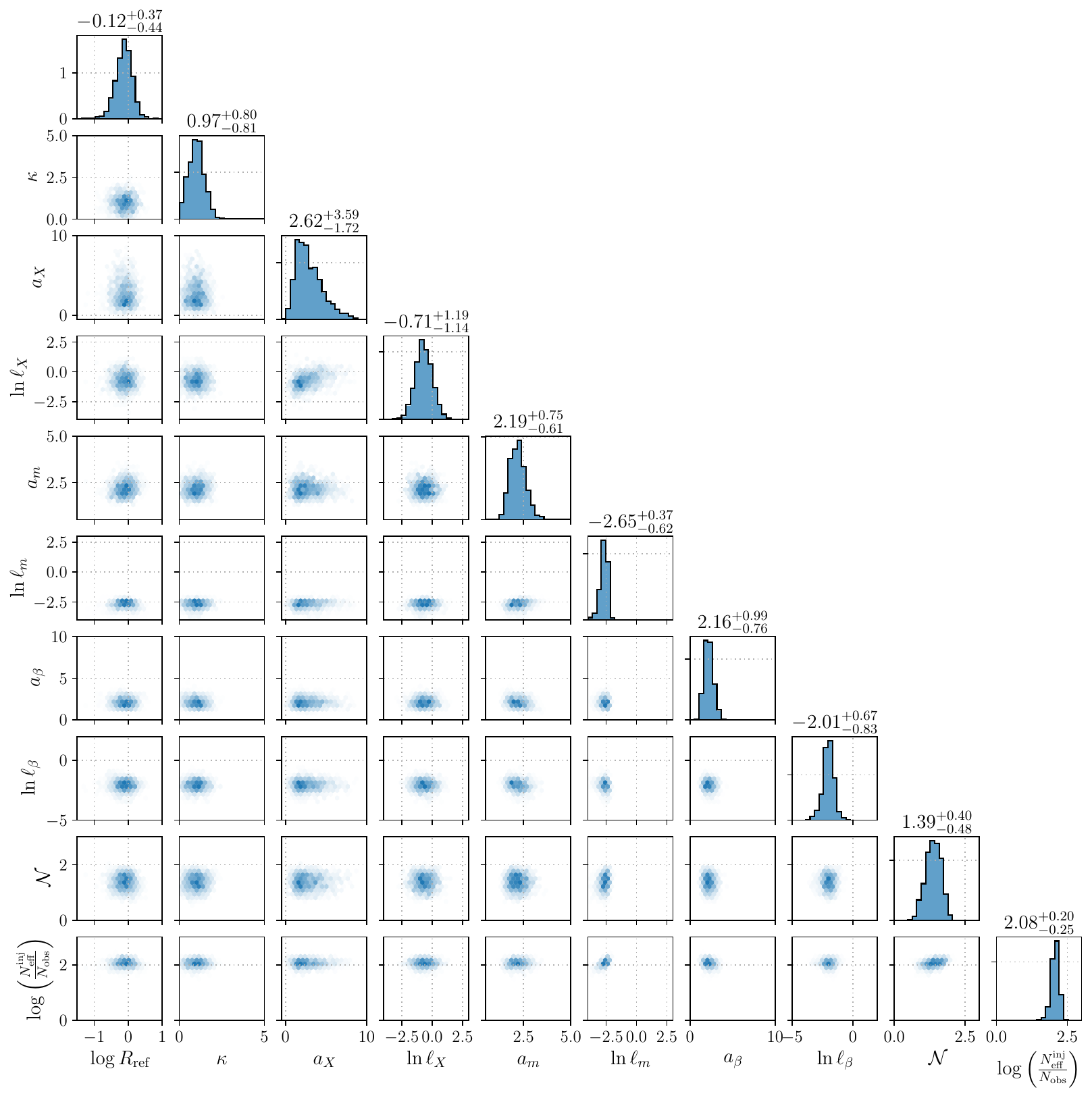}}
\caption{
Posterior distributions of hyper-parameters for the mass-interval population models of Section~\ref{sec:Methods}.
Here, $N_{\rm inj}^{\rm eff}/N_{\rm obs}$ gives the total number of injections
divided by the number of detections. The quantity $R_{\rm ref}$ represents the
differential merger-rate density, in units of
$\mathrm{Gpc}^{-3}\,\mathrm{yr}^{-1}\,M_\odot^{-1}$, evaluated at
$m_1 = 20\,M_\odot$ and redshift $z = 0.2$. In this model we set $(\tilde{m}_{\rm low},\,\tilde{m}_{\rm high})=
(18,30)\,M_{\odot}$.
}
\label{fig:param_posteriors}
\end{figure}

The prior choice for the hyper-parameters of our models are reported in Table~\ref{tab:priors}.
Figure~\ref{fig:param_posteriors} shows the recovered posteriors of key hyper-parameters for our models with $(\tilde{m}_{\rm low},\,\tilde{m}_{\rm high})=(50,~200)M_\odot$.

\begin{table}
\centering
\begin{tabular*}{\columnwidth}{@{\extracolsep{\fill}} l c c}
\hline
Parameter & Prior & Defined in \\ 
\hline
$a_{m_{1}}$ & $\mathcal{HN} (0.8)$ & $m_{1}$ Model \\
ln$\,l_{m_{1}}$ & $\mathcal{N} (-0.2, 1)$ & $m_{1}$ Model \\
\hline
$a_{\beta_{q}}$ & $\mathcal{HN} (1.0)$ & $\beta_{q}$ Model \\
ln$\,l_{\beta_{q}}$ & $\mathcal{N} (-0.5, 1))$ & $\beta_{q}$ Model \\
\hline
$a_{\chi_{eff}}$ & $\mathcal{HN} (3)$ & $\chi_{eff}$ Model \\
ln$\,l_{\chi_{eff}}$ & $\mathcal{N} (-0.5, 0.9)$ & $\chi_{eff}$ Model \\
\hline
$\chi_{\rm  max}$ & $\mathcal{U} [0.05,\, 1]$ & $\chi_{eff}$ Model  \\
$\chi_{\rm min,unscaled}$ & $\mathcal{U} [0,\, 1]$ &  $\chi_{eff}$ Model \\
\hline
$\mu$ & $\mathcal{U}(-1)$ & $p_{\rm out}(\chi_{\rm eff} \mid m_{1})$ Model  \\
$\sigma$ & $\mathcal{U}(-1.5, \, 0)$ & $p_{\rm out}(\chi_{\rm eff} \mid m_{1})$ Model  \\
 $\chi_{\rm max, out}$ & $\mathcal{U} [0.1,\, 1]$ & $p_{\rm out}(\chi_{\rm eff} \mid m_{1})$ Model \\ $\chi_{\rm min, unscaled, out}$ & $\mathcal{U} [0,\, 1]$ & $p_{\rm out}(\chi_{\rm eff} \mid m_{1})$ Model \\
\hline 
$\kappa$ & $\mathcal{N}(0, \, 5)$ & redshift Model \\
\hline
\end{tabular*}
\caption{Prior distributions for hyper-parameters of $p(\chi_{eff})$ and  mass models used in the main text. }
\label{tab:priors}
\end{table}

\end{document}